\documentclass[preprint,superscriptaddress,amsmath,amssymb,aps,prl]{revtex4-1}
\usepackage{graphicx,epstopdf}
\usepackage{dcolumn}
\usepackage{bm}
\usepackage{hyperref}
\usepackage{multirow}
\usepackage{threeparttable}
\usepackage{color}
\usepackage{subfigure}
\usepackage{setspace}
\usepackage{mathtools}
\usepackage{tipa}

\begin{document}


\title{Dynamic-Disorder-Induced Enhancement of Entanglement in Photonic Quantum Walks}

\author{Qin-Qin Wang}
\author{Xiao-Ye Xu}
\author{Wei-Wei Pan}
\author{Kai Sun}
\author{Jin-Shi Xu}
\author{Geng Chen}
\author{Yong-Jian Han}
\email{smhan@ustc.edu.cn}
\author{Chuan-Feng Li}
\email{cfli@ustc.edu.cn}
\author{Guang-Can Guo}
\affiliation{CAS Key Laboratory of Quantum Information, University of Science and Technology of China, Hefei 230026, People's Republic of China}
\affiliation{Synergetic Innovation Center of Quantum Information and Quantum Physics, University of Science and Technology of China, Hefei 230026, People's Republic of China}

\date{\today}

\begin{abstract}
Entanglement generation in discrete time quantum walks is deemed to be another key property beyond the transport behaviors. The latter has been widely used in investigating the localization or topology in quantum walks. However, there are few experiments involving the former for the challenges in full reconstruction of the final wave function. Here, we report an experiment demonstrating the enhancement of the entanglement in quantum walks using dynamic disorder. Through reconstructing the local spinor state for each site, von Neumann entropy can be obtained and used to quantify the coin-position entanglement. We find that the enhanced entanglement in the dynamically disordered quantum walks is independent of the initial state, which is different from the entanglement generation in the Hadamard quantum walks. Our results are inspirational for achieving quantum computing based on quantum walks.
\end{abstract}

\maketitle

\section{Introduction}

Entanglement, an intriguing character of quantum systems, plays the critical role in the quantum information processing\,\cite{Nielsen2000}, such as quantum key distribution\,\cite{Ekert1991} and quantum computing\,\cite{Shor1994}. However, quantum entanglement is so fragile that it can easily be destroyed by the noises and the environment. Systems become disordered\,\cite{Anderson1958} due to the inhomogeneous environmental conditions and other parameters in the system, which are impossible to control experimentally. Intuitively, one may expect that such disorder would reduce the entanglement of a given system, which is indeed true for a large variety of systems. In fact for some systems, disorder can enhance the entanglement\,\cite{Niederberger2010,Prabhu2011,Mishra2016,Vieira2013,Vieira2014,Zeng2017}. For example, the genuine multipartite entanglement of ground state in the quantum spin model\,\cite{Binosi2007,Prabhu2011} can be enhanced by disorder. Another example is the dynamic disorder that can enhance the entanglement generation in quantum walks (QW)\,\cite{Vieira2013,Vieira2014,Zeng2017}.

\begin{figure}
  \centering
  \includegraphics[width=0.40\textwidth]{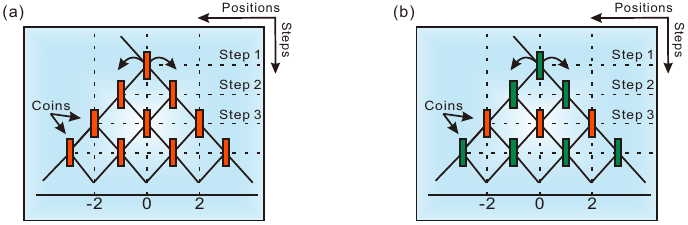}\\
  \includegraphics[width=0.47\textwidth]{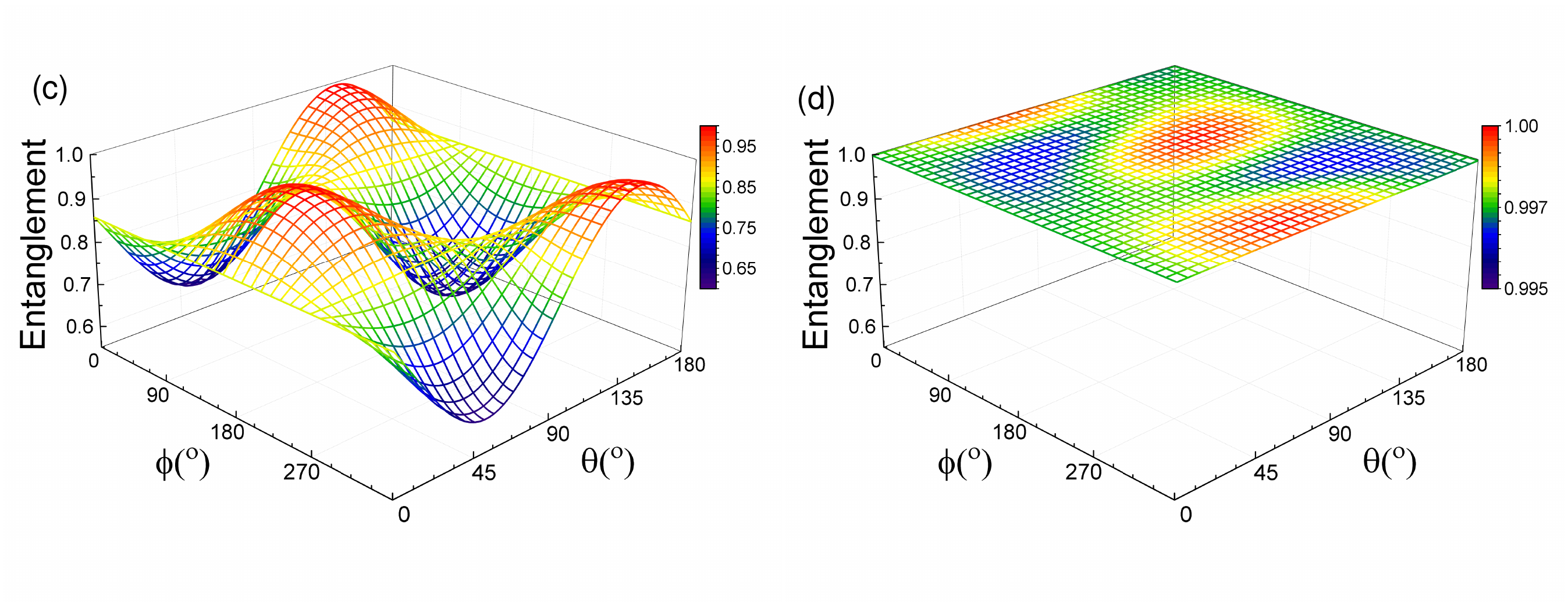}
  \caption{Illustrations for (a) ordered QW, where only one kind of coin is employed (red rectangle), and for (b) dynamically disordered QW, where two kinds of coin are employed (red and green rectangles). The coin-position entanglement as a function of the initial state parameters $\{\theta,\phi\}$ for (c) ordered Hadamard QW and for (d) the dynamically disordered QW with the coin tossing  $S^0_C$ (define in the text). The dependency of the coin-position entanglement on the initial state is significantly higher for ordered QW compared to the dynamically disordered one.}\label{fig.entropy}
\end{figure}

QW on lattices and graphs\,\cite{Aharonov1993} is a quantum generalization of the classical random walks\,\cite{Karl1905}, which may play as universal quantum computers\,\cite{Childs2009,Childs2013},  quantum simulators\,\cite{Chandrashekar2013} and platforms to investigate topological phases\,\cite{Kitagawa2010,Kitagawa2012,rechtsman2016}. The behavior of QW, especially the ballistic behavior\,\cite{Blanchard2004} of the transport properties, is dramatically different from its classical counterpart due to the superposition principle and has been extensively investigated. Besides the probability distribution, the entanglement properties of QW have also been theoretically studied\,\cite{Mackay2002,Carneiro2005,Franco2011,Venegas2012}. It is the genuine quantum feature in QW since there is no classical counterpart. 
The entanglement (coin-position entanglement) here is different from its original definition for multiple parties. It is actually entanglement between two modes sharing a single particle\,\cite{Morin2013}, which has been widely used in some crucial quantum information protocols\,\cite{Choi2008}. In the ordered QW where the quantum coin is fixed during the whole evolution process or changes with a deterministic way, the coin-position entanglement\,\cite{Carneiro2005,Mackay2002,Franco2011} is highly dependent on the initial state and usually never achieves its maximal value. Entanglement in QW will be affected by dynamic disorder, in which the quantum coin is independent of the site $j$ and at the same time is randomly chosen for each step. Or it can be affected by the static disorder, in which the quantum coin is fixed for all the time and at the same time is randomly chosen for each site $j$. Entanglement in QW can also be affected by the combination of both, dynamic and static disorder. Theoretical investigations\,\cite{Vieira2013,Vieira2014} showed that the entanglement is reduced by the static disorder while the dynamic disorder induces enhancement of the entanglement independent of the initial states and even with the appearance of the static disorder.

Linear optics is a good platform for implementing QW and thus many technologies have been developed: spatial displacers\,\cite{Kitagawa2012}, orbital angular momentum (OAM)\,\cite{Cardano2015}, time multiplexing\,\cite{Schreiber2010,Schreiber2011}, integrated optical circuits\,\cite{Sansoni2012} and array of wave guides\,\cite{Peruzzo2010,titchener2016}. Unlike transport behaviors, which have been sufficiently studied just by measuring the final probability distribution, the entanglement properties are still needed to be studied in both ordered and disordered QW. The experimental challenges are twofold: how to reach large-scale QW (the disorder-induced entanglement enhancement can only be demonstrated in the asymptotic limit) and how to reconstruct the wave function in both the coin and position space\,\cite{Franco2011}. Different efforts\,\cite{Perets2008,Regensburger2011,Schreiber2012} have been made to improve the scalability. For example, all fiber-based QW have reached 62 steps with high fidelity and low loss\,\cite{Boutari2016}. Only recently, the final wave function in a one-dimensional QW of a single cesium (Cs) atom has been obtained by the local quantum state tomography\,\cite{Karski2009}, and complete reconstruction of wave function was achieved in OAM\,\cite{Cardano2015} and time multiplexing protocol\,\cite{Xu2018} (in Ref.\,\cite{Barkhofen2017} the authors measured the relative phase, $0$ or $\pi$, between the neighboring sites). In this paper,we report an experiment for demonstrating the enhancement of entanglement generation in QW by the dynamic disorder. This experiment is based on our recently developed novel compact platform for genuine single photon QW in large scale with the ability of full wave function reconstruction.

\section{Theoretical idea}

\begin{figure}
  \centering
  \includegraphics[width=0.47\textwidth]{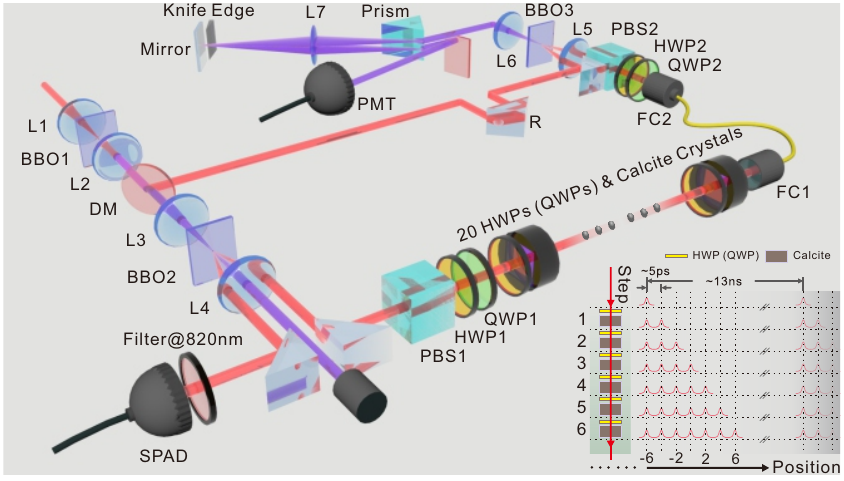}
  \caption{Diagram of the experimental setup (additional details are given in the supplementary material). The system contains four parts: 1. Second harmonic generation in BBO1 to obtain the ultraviolet pulse; 2. Generation of heralded single photons by adopting the beam-like type SPDC in BBO2; 3. Time-multiplexing quantum walks realized by birefringent crystals sketched by the inset at the bottom right corner; 4. Ultra-fast detection of the arrival time of single photons with an up-conversion single photon detector. BBO: $\beta$-BaB$_2$O$_4$, L: lens, DM: dichroic mirror, PBS: polarization dependent beam splitter, HWP: half-wave plate, QWP: quarter-wave plate, R: reflector, FC: fiber collimator, PMT: photomultiplier tubes, SPAD: single-photon avalanche diode.}\label{fig.setup}
\end{figure}

The Hilbert space of a QW is $H = H_C\otimes H_P$, where $H_C$ is a two-dimensional Hilbert space spanned by $\{|{\uparrow}\rangle,|{\downarrow}\rangle\}$ and $H_P$ is an infinite dimensional Hilbert space spanned by a set of orthogonal vectors $|{j}\rangle(j\in \mathbb{Z})$. A QW is given by a sequence of coin tossing followed by a conditional shift according to the coin state. The time evolution operator for a QW from $t = 0$ to $T$ can be represented by a multi-step unitary operator $\mathcal{U}(T) \equiv \prod_{t = 0}^{T}U(t)$, where $U(t) = S\cdot(C(t)\otimes\mathbb{I}_P)$ with $\mathbb{I}_P$ is the identity operator in $H_P$ and $C(t)$ is the coin tossing in $H_C$. The shift operator $S = \sum_{j}|{\uparrow}\rangle\langle{\uparrow}|\otimes|{j+1}\rangle\langle{j}| + |{\downarrow}\rangle\langle{\downarrow}|\otimes|{j-1}\rangle\langle{j}|$ describes the conditional displacement in lattice, which generates the coin-position entanglement.

Generally, the coin tossing $C(t)$ in a QW is time and position dependent. In this paper, $C(t)$ is assumed to be site independent since we only consider the effect of the dynamic disorder. In a QW with dynamic disorder, the coin tossing $C(t)$ is time dependent: for each step, it is randomly chosen from a set $\mathbb{C}$ with certain probability distribution (in particularly, homogeneous distribution). According to the literature\,\cite{Vieira2013,Vieira2014}, the type of dynamic disorder (including type of $\mathbb{C}$) is not important. Without loss of generality, in our experiment, we assumed that $\mathbb{C}$ consists of Hadamard operator ($H$) and Fourier operator ($F$), where
\begin{equation}\label{operation}
   H = \frac{1}{\sqrt{2}}\left (
   \begin{array}{cc}
   1 & 1 \\
   1 & -1
   \end{array}\right ),~~~
   F = \frac{1}{\sqrt{2}}\left (
   \begin{array}{cc}
   1 & i \\
   i & 1
   \end{array}\right ).
\end{equation}
We also considered an ordered QW, in which the coin tossing is time independent, and we chose Hadamard gate all the time for comparison.

The global time evolution operator for a single sample is also unitary in a  QW and the final state $|{\Psi(t)}\rangle$ after a $t$-step walk remains pure if the initial state is pure. The general form of $|{\Psi(t)}\rangle$ can be written as $\sum_{j}[a(j,t)|{\uparrow}\rangle+b(j,t)|{\downarrow}\rangle]\otimes|{j}\rangle$, where $a(j,t)$ and $b(j,t)$ are complex numbers with the normalization condition $\sum_{j}[|a(j,t)|^2 + |b(j,t)|^2] = 1$. With the unitarity factor, the coin-position entanglement in a QW can be defined by the von Neumann entropy
\begin{equation}\label{vonEntropy}
  S_E(\rho(t)) = -\text{Tr}[\rho_C(t)\log_2\rho_C(t)],
\end{equation}
where $\rho_C(t) = \text{Tr}_P[\rho(t)]$ is the reduced density matrix of the coin with $\rho(t) = |{\Psi(t)}\rangle\langle{\Psi(t)}|$ and $\text{Tr}_P$ is taking the trace over position.

For a fixed initial state, with the increase in time, regardless of the ordered or disordered QW, the coin-position entanglement will be asymptotic to a stable value. Generally, this asymptotic value in an ordered QW can not reach maximal and is strongly dependent on the initial state. The entanglement after 20 steps with different initial states is shown in Fig.\,\ref{fig.entropy}.(c). For a dynamically disordered QW, this asymptotical value is found\,\cite{Vieira2013,Vieira2014} to reach maximal regardless of the initial states as shown in Fig.\,\ref{fig.entropy}.(d).

\section{Experimental setup and results}

\begin{figure}[t]
  \centering
  \includegraphics[width=0.45\textwidth]{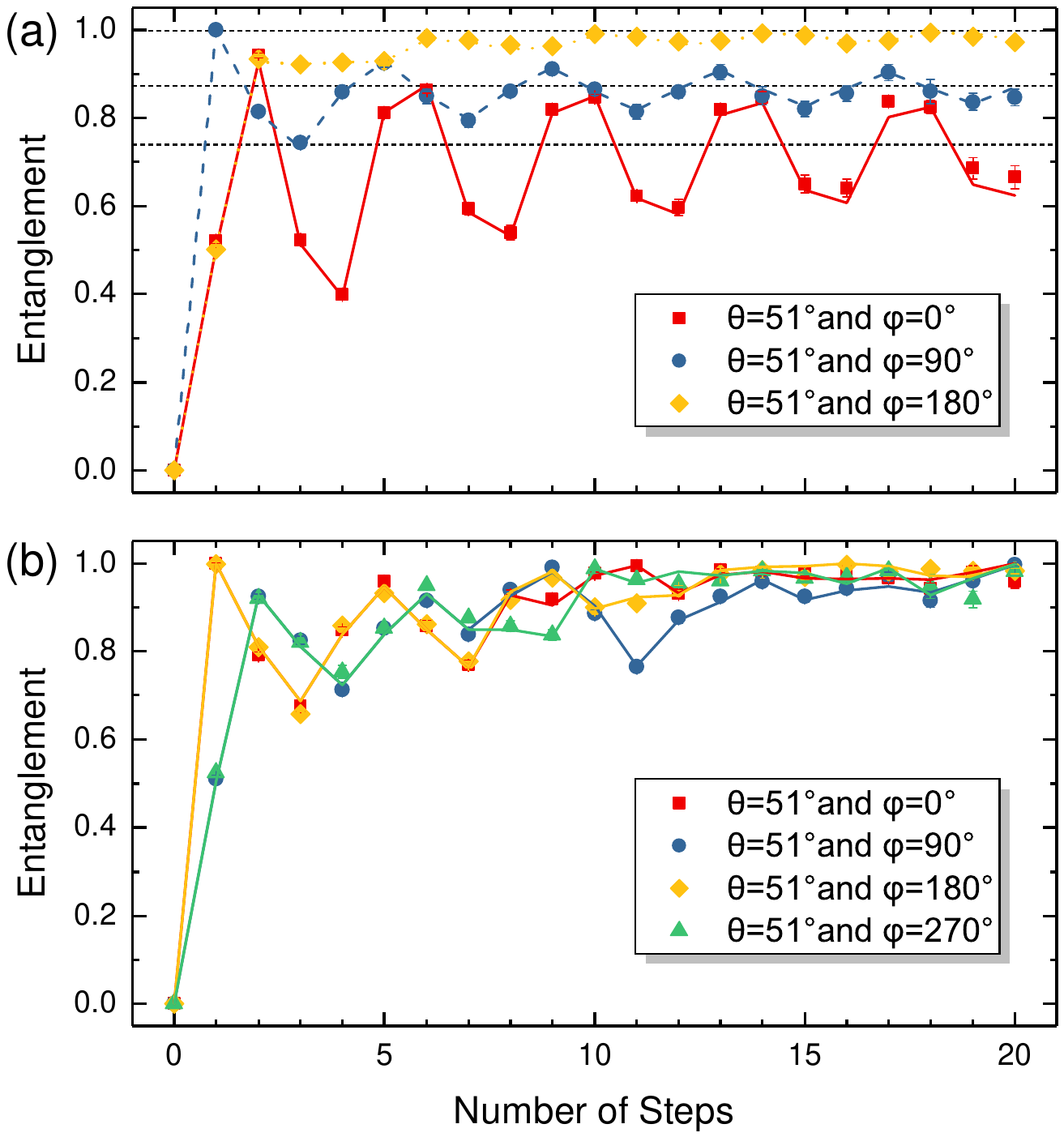}
  \caption{(a) Coin-position entanglement measured (dots) for different initial states in ordered Hadamard QW. The lines (solid red, dashed blue and dotted yellow) are the theoretical predictions of the von Neumann entropies. The entanglement generated in ordered QW shows a significant dependency on the initial state. Noting that the entanglement involves periodical oscillation around an asymptotic value (guided by the dashed black horizontal line) in the limit of infinite steps with an amplitude decay. We have not considered the case $\{51^\circ, 270^\circ\}$, which is the same as $\{51^\circ, 90^\circ\}$. (b) Coin-position entanglement measured (dots) for different initial states in the dynamically disordered scenario. The lines (solid red, solid blue, solid yellow and solid green) are the theoretical predictions. The von Neumann entropies are about 0.98 at 20 steps for all initial state clusters, which significantly differ from the ordered case, where they converge to different asymptotic values for different initial states. The periodical oscillation of entanglement with time degenerates. Only statistical errors are considered with total counts 24 thousand in 4 hundred seconds.}\label{fig.EnResults}
\end{figure}

The experimental setup is shown in Fig.\,\ref{fig.setup} and the more detail description is given in supplementary material. Single photons generated from spontaneous parametric down conversion(SPDC) are adopted as the herald walker. These kind of coin states are initialized by sending them through the polarizer PBS1-HWP1-QWP1(see Fig.\,\ref{fig.setup}). The state $|{\uparrow}\rangle(|{\downarrow}\rangle)$ corresponds to a single horizontally polarized photon $|{H}\rangle(|{V}\rangle)$, which stands for the horizontal (vertical) polarization of the photon (walker). The QW device is composed of wave plates (for realizing coin tossing) and calcite crystals (for implementing conditional shift), and each step contains each one of them. $H$ and $F$ coin tossing was implemented by single HWP (with its optical axis oriented at $\pi/8$) and QWP (with its optical axis oriented at $-\pi/4$) respectively. The reduced density matrix $\rho_C$ in Eq.\,\ref{vonEntropy} is equal to $\sum_{j}p_j\rho_j$. Equation $p_j = |a(j)|^2 + |b(j)|^2$ is the probability in site $j$, $\rho_j$ is the local density matrix in site $j$. In experiment, $\rho_j$ is obtained through local quantum state tomography (realized by the polarization analyzer QWP2-HWP2-PBS2 in Fig.\,\ref{fig.setup}). Meanwhile $p_j$ is directly given by the projection probabilities in the first two bases, $|{H}\rangle\langle{H}|$ and $|{V}\rangle\langle{V}|$. The lattice is composed of arriving time of signal photons and the time interval is around $5$\,$ps$, which is challenging to detect with available commercial detectors. Therefore we constructed the up-conversion single photon detectors.

\begin{figure}[t]
  \centering
  \includegraphics[height=53mm,width=0.47\textwidth]{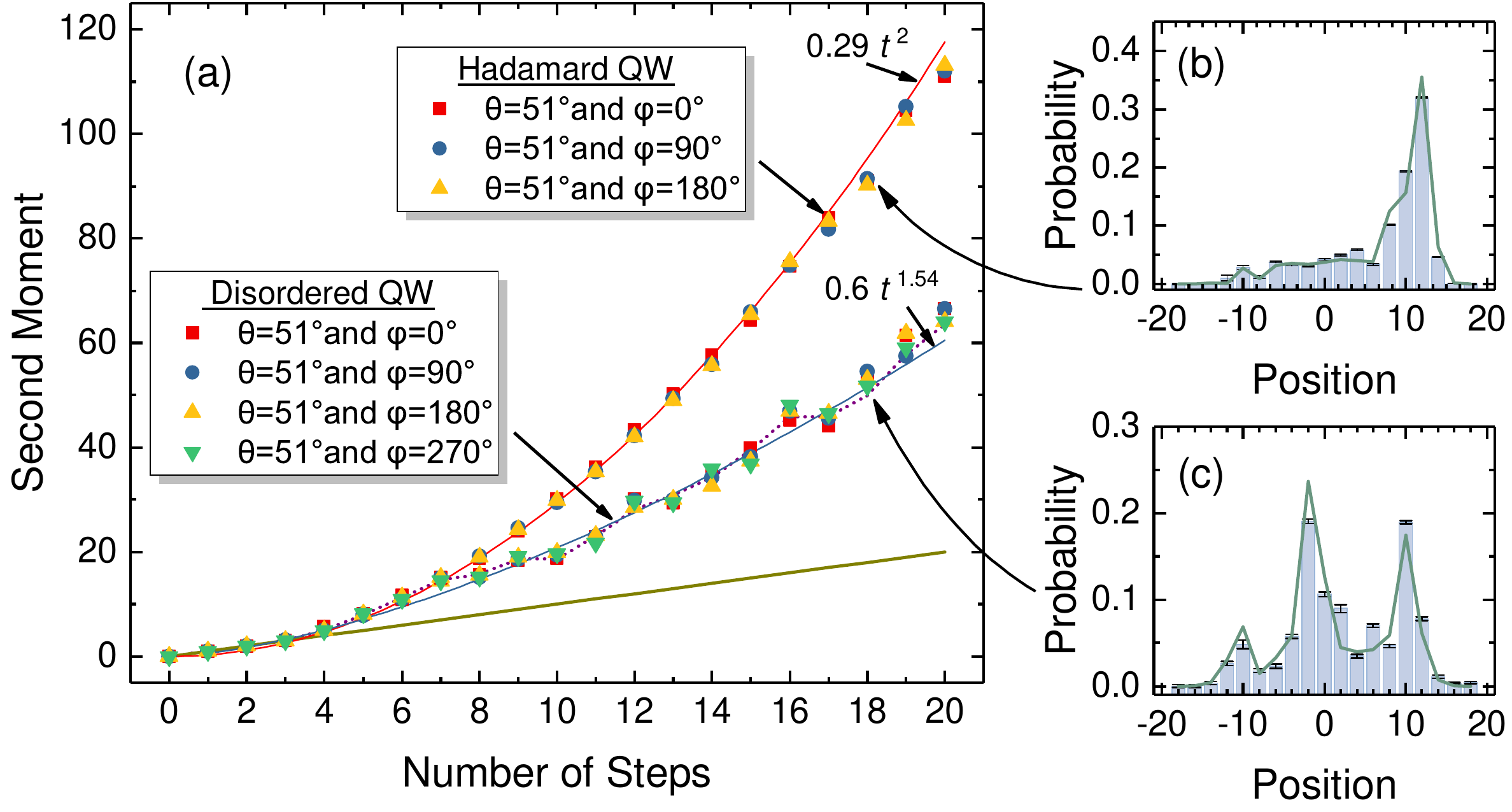}
  \caption{Measured trend (a) of second moment up to 20 steps (dots) for different initial states in ordered and dynamically disordered QW. Ballistic transport behavior is observed in ordered QW (solid red line, fitted as $0.29t^2$). In dynamically disordered QW, a sub-ballistic transport behavior is observed (solid blue line, fitted as $0.6t^{1.54}$) and the dotted red line gives the theoretical predictions. The deep yellow line represents the typical diffuse transport behavior in classical random walks starting at the origin. The error bars are smaller than the point size with only consideration of statistical errors. (b) and (c) give the probability distributions after a 18-step QW for initial state with $\theta = 51^\circ$ and $\phi = 0^\circ$. It is seen clearly that the spreading velocity in ordered QW (b) is significantly faster than dynamically disordered one (c).}\label{fig.Var}
\end{figure}

The initial state in our experiments is located at the original site ($|x=0\rangle$) and the general state of the coin is $|{\Psi(0)}\rangle = [\cos(\theta/2)|{\uparrow}\rangle + e^{i\phi}\sin(\theta/2)|{\downarrow}\rangle]\otimes|{0}\rangle$, where $\theta\in[0,\pi]$ and $\phi\in[0, 2\pi)$. In our experiment, the QW step number is limited to 20. First, we experimentally determined the key characteristics of the coin-position entanglement generated in the standard Hadamard QW (ordered QW). We chose three different initial states: $\{\theta,\phi\}$=$\{51^\circ, 0^\circ\}$, $\{51^\circ, 90^\circ\}$ and $\{51^\circ, 180^\circ\}$. The entanglement dynamics were experimentally obtained for each initial state (Fig.\,\ref{fig.EnResults}(a)). Theoretically, the entanglement for a given initial state will approach an asymptotic value after several oscillations. In addition, the asymptotic value is strongly dependent on the initial state: $S_E = 0.739$ for $\phi = 0^\circ$, $S_E=0.867$ for $\phi = 90^\circ$ and $S_E = 0.977$ for $\phi = 180^\circ$ (dashed guided lines in Fig.\,\ref{fig.EnResults}(a)). In our experiment, the entanglement almost approached the theoretical asymptotic vale for $\phi=90^\circ$ and $\phi=180^\circ$. At $\phi = 0^\circ$, the experimental result was $0.665\pm0.027$ after a 20-step QW and the entanglement was still oscillating. Besides, the ballistic transport behavior of ordered QW is shown in Fig.\,\ref{fig.Var}(a). During the experiment, the fidelities, defined as $F(\rho_C^{\text{exp}}, \rho_C^{\text{th}}) = \text{Tr}[\sqrt{\sqrt{\rho_C^{\text{exp}}}\rho_C^{\text{th}}\sqrt{\rho_C^{\text{exp}}}}]^2$ with $\rho^\text{exp(th)}$ representing the experimentally measured (theoretically predicted) density matrix, are larger than $0.986 \pm 0.001$ for each initial state and step.

We further demonstrated that the dynamic disorder can enhance the coin-position entanglement. To achieve this, we first chose the initial coin state as $\{\theta,\phi\} = \{51^\circ, 0^\circ\}$, where the coin-position entanglement after a 20-step Hadamard QW is minimal (see Fig.\,\ref{fig.entropy}(c) and Fig.\,\ref{fig.EnResults}(a)). We showed that the coin-position entanglement can be dramatically enhanced to about $S_E=0.98$ by the dynamically disordered coin tossing sequence $S_C^0 = FFHFHFHHFFFFFHFHHHHH$ (operated on the coin from left to right). Actually, the sequence $S_C^0$ is one of the optimal sequences to enhance the entanglement for the initial state $\{51^\circ, 0^\circ\}$ after 20 steps. $S_C^0$ can also enhance the entanglement for any of the initial states (the theoretical enhanced entanglement with the sequence $S_C^0$ for any initial state can be found in Fig.\,\ref{fig.entropy}(d)). We checked the enhancement with three other initial states: $\{51^\circ, 90^\circ\}$, $\{51^\circ, 180^\circ\}$ and $\{51^\circ, 270^\circ\}$. The experimental results are shown in Fig.\,\ref{fig.EnResults}(b), which clearly shows that all the entanglements are improved and approach the asymptotic value faster than in the ordered scenario. More importantly, the entanglement approached the same value (about $0.98$), which is almost equal to the theoretical maximal value of $1$ regardless of the initial state. The fidelities are larger than $0.968 \pm 0.003$ for each initial state and step in this scenario. Actually, the wave-function transport in a lattice will decelerate and show a sub-ballistic behavior in the presence of disorder\,\cite{Eichelkraut2013}. The dynamic disorder can lead to a sub-ballistic transport behavior in a QW, and the transport trend depends on the choice of the two coin operations and the sequences \,\cite{Vieira2013,Ribeiro2004}. In Fig.\,\ref{fig.Var}(a), we show the sub-ballistic transport behavior in a dynamically disordered QW. It is obvious that its spreading velocity is faster than a typical diffuse transport behavior in a classical random walk but slower than a ballistic transport behavior in an ordered QW.

\begin{figure}
  \centering
  \includegraphics[height=53mm,width=0.47\textwidth]{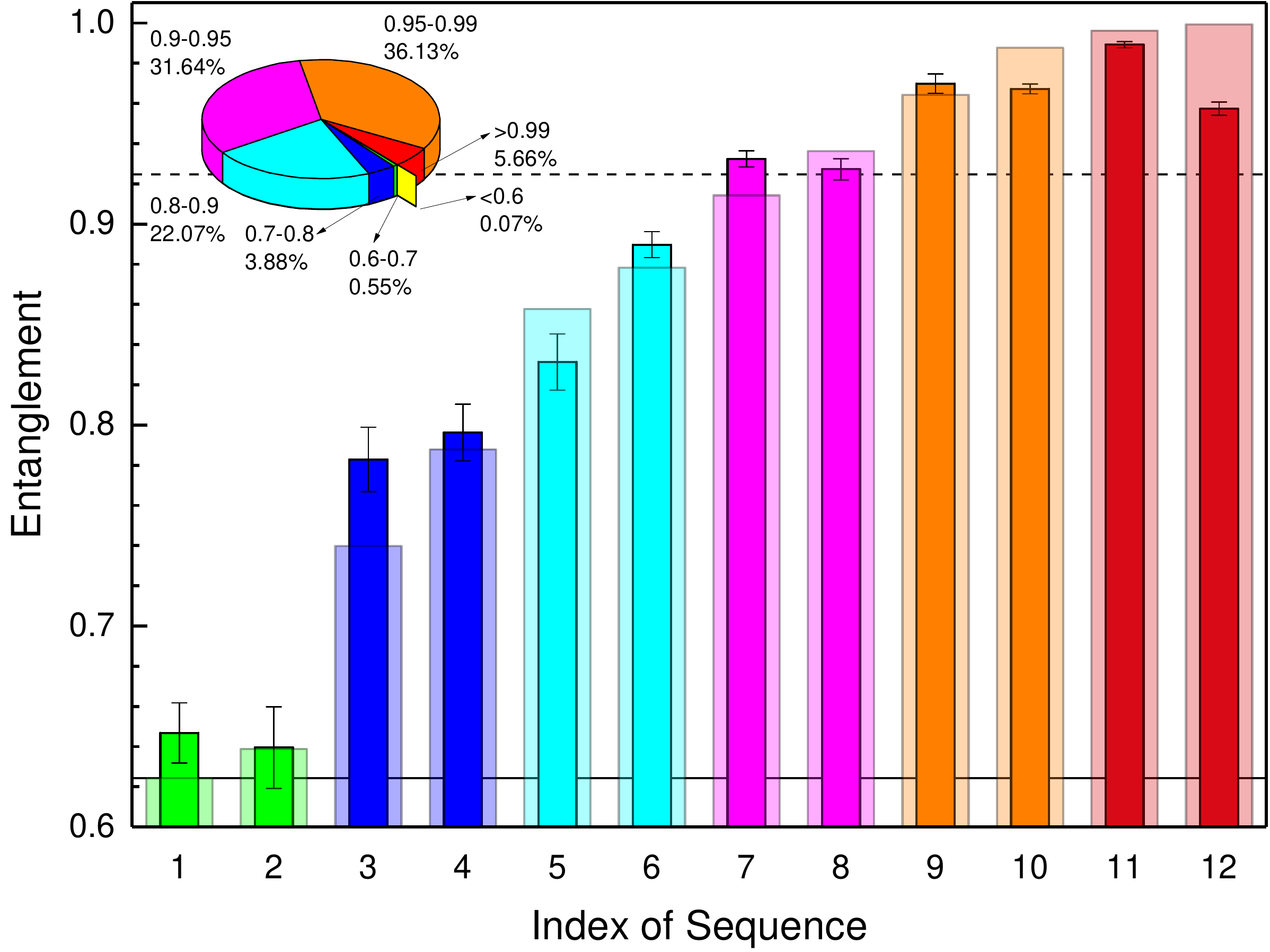}
  \caption{Coin-position entanglement measured (opaque bars) for different sequences (forms are given in supplementary material) in dynamically disordered QW with the theoretical predictions shown by transparent bars. The inset shows the rate of sequences generating different $S_E\in\{0,1\}$ at 20 steps with the initial state $\{51^\circ, 0^\circ\}$, whose entanglement is lowest (shown with the solid black line) in ordered Hadamard QW. Note that almost 73\% of sequences can generate $S_E>0.9$. In each entanglement interval (different color), we randomly choose two sequences. The theoretical average entanglement $\langle S_E\rangle = 0.924$ (guided by the black-dash horizontal line) was calculated by averaging all sequences. Only statistical errors are considered.}\label{fig.Disorder}
\end{figure}

Theoretically, the enhancement of coin-position entanglement is not dependent on the specific form of the coin tossing sequence when randomness is introduced, and the number of the steps is infinite. However, in our experiment, the total amount of step was limited to 20. In this case, the enhancement of entanglement is dependent on the sequence $S_C$. The dependence of the final entanglement after a 20-step disordered QW with the initial state $\{51^\circ, 0^\circ\}$ on the disordered sequence $S_C$ is shown in the Fig.\,\ref{fig.Disorder}. Based on this dependence, the sequences can be divided into several different types (indicated in different colors in Fig.\,\ref{fig.Disorder}). We experimentally checked the enhancement of the entanglement after 20-step QW with different types of sequence (two sequences in each type), and the experimental results are shown in Fig.\,\ref{fig.Disorder}. The disordered QW became more powerful than the ordered ones in terms of ability to generate entanglement. If we have no information about the sequence, the entanglement should be obtained by averaging all the sequences. Based on our experimental results, the average entanglement $\langle S^{\text{exp}}_E\rangle$, defined as $\langle S^{\text{exp}}_{E}\rangle=\sum_{i=1}^{12}S^{i}_{E}\cdot \frac{P_{i}}{2}$ with $P_{i}$ being the rate of entanglement interval, to which $S^{i}_{E}$ belongs, is about $0.923\pm0.006$, which is a significant improvement for the ordered QW.

\section{Conclusions}

In conclusion, we reported the first experiment to study the coin-position entanglement generation in discrete time quantum walks beyond the usual transport behaviors. We observed the initial state dependency of entanglement generation in ordered quantum walks. This entanglement involves periodic oscillations with the amplitude decay around an asymptotic value, which is dependent on the initial state. More importantly, we found the coin-position entanglement can be significantly improved by the dynamic disorder for any initial state. Besides, we showed the sub-ballistic transport behavior in dynamically disordered quantum walks. Based on our experimental results, it seems that the entanglement power of the coin tossing sequence $S_C$ positively correlates with the complexity of the sequence. In the spirit of the Kolmogorov complexity, the complexity of a binary sequence can be explicitly defined via Lempel-Ziv complexity $C_{\text{LZ}}$\,\cite{Lempel1976,Crutchfield2011}. The complexity of the twelve sequences (from 1 to 12) in the Fig.\,\ref{fig.Disorder} are $3$, $3$, $5$, $8$, $7$, $7$, $6$, $6$, $6$, $7$, $7$ and $6$, respectively (details given in the supplementary). The complexity measure is qualitative, and our results qualitatively showed that the power of entanglement generation increased with the complexity of the sequence. When the sequence length increased, the complexity of random sequence also increased and the entanglement power of the sequence increased as a result as well. However, the complexity of periodic sequence will be eventually saturated, and the entanglement power will not increase. When the disordered sequence approached infinity, the complexity will be infinite and the entanglement power of all the disordered sequences will be the same and will be a maximal entanglement generator. Our experiment applies a way to explore the entanglement in quantum information.

\begin{acknowledgments}
Qin-Qin Wang and Xiao-Ye Xu contributed equally to this work. This work was supported by National Key Research and Development Program of China (Nos.\,2017YFA0304100, 2016YFA0302700), the National Natural Science Foundation of China (Nos.\,61327901, 11474267, 11774335, 61322506), Key Research Program of Frontier Sciences, CAS (No.\,QYZDY-SSW-SLH003), the Fundamental Research Funds for the Central Universities (No.\,WK2470000026), the National Postdoctoral Program for Innovative Talents (No.\,BX201600146), China Postdoctoral Science Foundation (No.\,2017M612073), and Anhui Initiative in Quantum Information Technologies (Grant No. AHY060300).
\end{acknowledgments}

\bibliographystyle{apsrev4-1.bst}
\bibliography{reference}

\begin{thebibliography}{44}%
\makeatletter
\providecommand \@ifxundefined [1]{%
 \@ifx{#1\undefined}
}%
\providecommand \@ifnum [1]{%
 \ifnum #1\expandafter \@firstoftwo
 \else \expandafter \@secondoftwo
 \fi
}%
\providecommand \@ifx [1]{%
 \ifx #1\expandafter \@firstoftwo
 \else \expandafter \@secondoftwo
 \fi
}%
\providecommand \natexlab [1]{#1}%
\providecommand \enquote  [1]{``#1''}%
\providecommand \bibnamefont  [1]{#1}%
\providecommand \bibfnamefont [1]{#1}%
\providecommand \citenamefont [1]{#1}%
\providecommand \href@noop [0]{\@secondoftwo}%
\providecommand \href [0]{\begingroup \@sanitize@url \@href}%
\providecommand \@href[1]{\@@startlink{#1}\@@href}%
\providecommand \@@href[1]{\endgroup#1\@@endlink}%
\providecommand \@sanitize@url [0]{\catcode `\\12\catcode `\$12\catcode
  `\&12\catcode `\#12\catcode `\^12\catcode `\_12\catcode `\%12\relax}%
\providecommand \@@startlink[1]{}%
\providecommand \@@endlink[0]{}%
\providecommand \url  [0]{\begingroup\@sanitize@url \@url }%
\providecommand \@url [1]{\endgroup\@href {#1}{\urlprefix }}%
\providecommand \urlprefix  [0]{URL }%
\providecommand \Eprint [0]{\href }%
\providecommand \doibase [0]{http://dx.doi.org/}%
\providecommand \selectlanguage [0]{\@gobble}%
\providecommand \bibinfo  [0]{\@secondoftwo}%
\providecommand \bibfield  [0]{\@secondoftwo}%
\providecommand \translation [1]{[#1]}%
\providecommand \BibitemOpen [0]{}%
\providecommand \bibitemStop [0]{}%
\providecommand \bibitemNoStop [0]{.\EOS\space}%
\providecommand \EOS [0]{\spacefactor3000\relax}%
\providecommand \BibitemShut  [1]{\csname bibitem#1\endcsname}%
\let\auto@bib@innerbib\@empty
\bibitem [{\citenamefont {Nielsen}\ and\ \citenamefont
  {Chuang}(2000)}]{Nielsen2000}%
  \BibitemOpen
  \bibfield  {author} {\bibinfo {author} {\bibfnamefont {M.~A.}\ \bibnamefont
  {Nielsen}}\ and\ \bibinfo {author} {\bibfnamefont {I.~L.}\ \bibnamefont
  {Chuang}},\ }\href@noop {} {\emph {\bibinfo {title} {Quantum computation and
  quantum information}}}\ (\bibinfo  {publisher} {Cambridge University Press},\
  \bibinfo {address} {Cambridge ; New York},\ \bibinfo {year} {2000})\ pp.\
  \bibinfo {pages} {xxv, 676 p.}\BibitemShut {Stop}%
\bibitem [{\citenamefont {Ekert}(1991)}]{Ekert1991}%
  \BibitemOpen
  \bibfield  {author} {\bibinfo {author} {\bibfnamefont {A.~K.}\ \bibnamefont
  {Ekert}},\ }\href {\doibase 10.1103/PhysRevLett.67.661} {\bibfield  {journal}
  {\bibinfo  {journal} {Phys. Rev. Lett.}\ }\textbf {\bibinfo {volume} {67}},\
  \bibinfo {pages} {661} (\bibinfo {year} {1991})}\BibitemShut {NoStop}%
\bibitem [{\citenamefont {Shor}(1994)}]{Shor1994}%
  \BibitemOpen
  \bibfield  {author} {\bibinfo {author} {\bibfnamefont {P.~W.}\ \bibnamefont
  {Shor}},\ }in\ \href {\doibase 10.1109/SFCS.1994.365700} {\emph {\bibinfo
  {booktitle} {Foundations of Computer Science, 1994 Proceedings., 35th Annual
  Symposium on}}}\ (\bibinfo {organization} {Ieee},\ \bibinfo {year} {1994})\
  pp.\ \bibinfo {pages} {124--134}\BibitemShut {NoStop}%
\bibitem [{\citenamefont {Anderson}(1958)}]{Anderson1958}%
  \BibitemOpen
  \bibfield  {author} {\bibinfo {author} {\bibfnamefont {P.~W.}\ \bibnamefont
  {Anderson}},\ }\href {\doibase 10.1103/PhysRev.109.1492} {\bibfield
  {journal} {\bibinfo  {journal} {Phys. Rev.}\ }\textbf {\bibinfo {volume}
  {109}},\ \bibinfo {pages} {1492} (\bibinfo {year} {1958})}\BibitemShut
  {NoStop}%
\bibitem [{\citenamefont {Niederberger}\ \emph {et~al.}(2010)\citenamefont
  {Niederberger}, \citenamefont {Rams}, \citenamefont {Dziarmaga},
  \citenamefont {Cucchietti}, \citenamefont {Wehr},\ and\ \citenamefont
  {Lewenstein}}]{Niederberger2010}%
  \BibitemOpen
  \bibfield  {author} {\bibinfo {author} {\bibfnamefont {A.}~\bibnamefont
  {Niederberger}}, \bibinfo {author} {\bibfnamefont {M.~M.}\ \bibnamefont
  {Rams}}, \bibinfo {author} {\bibfnamefont {J.}~\bibnamefont {Dziarmaga}},
  \bibinfo {author} {\bibfnamefont {F.~M.}\ \bibnamefont {Cucchietti}},
  \bibinfo {author} {\bibfnamefont {J.}~\bibnamefont {Wehr}}, \ and\ \bibinfo
  {author} {\bibfnamefont {M.}~\bibnamefont {Lewenstein}},\ }\href {\doibase
  10.1103/PhysRevA.82.013630} {\bibfield  {journal} {\bibinfo  {journal} {Phys.
  Rev. A}\ }\textbf {\bibinfo {volume} {82}},\ \bibinfo {pages} {013630}
  (\bibinfo {year} {2010})}\BibitemShut {NoStop}%
\bibitem [{\citenamefont {Prabhu}\ \emph {et~al.}(2011)\citenamefont {Prabhu},
  \citenamefont {Pradhan}, \citenamefont {Sen(De)},\ and\ \citenamefont
  {Sen}}]{Prabhu2011}%
  \BibitemOpen
  \bibfield  {author} {\bibinfo {author} {\bibfnamefont {R.}~\bibnamefont
  {Prabhu}}, \bibinfo {author} {\bibfnamefont {S.}~\bibnamefont {Pradhan}},
  \bibinfo {author} {\bibfnamefont {A.}~\bibnamefont {Sen(De)}}, \ and\
  \bibinfo {author} {\bibfnamefont {U.}~\bibnamefont {Sen}},\ }\href {\doibase
  10.1103/PhysRevA.84.042334} {\bibfield  {journal} {\bibinfo  {journal} {Phys.
  Rev. A}\ }\textbf {\bibinfo {volume} {84}},\ \bibinfo {pages} {042334}
  (\bibinfo {year} {2011})}\BibitemShut {NoStop}%
\bibitem [{\citenamefont {Mishra}\ \emph {et~al.}(2016)\citenamefont {Mishra},
  \citenamefont {Rakshit}, \citenamefont {Prabhu}, \citenamefont {De},\ and\
  \citenamefont {Sen}}]{Mishra2016}%
  \BibitemOpen
  \bibfield  {author} {\bibinfo {author} {\bibfnamefont {U.}~\bibnamefont
  {Mishra}}, \bibinfo {author} {\bibfnamefont {D.}~\bibnamefont {Rakshit}},
  \bibinfo {author} {\bibfnamefont {R.}~\bibnamefont {Prabhu}}, \bibinfo
  {author} {\bibfnamefont {A.~S.}\ \bibnamefont {De}}, \ and\ \bibinfo {author}
  {\bibfnamefont {U.}~\bibnamefont {Sen}},\ }\href {\doibase
  10.1088/1367-2630/18/8/083044} {\bibfield  {journal} {\bibinfo  {journal}
  {New J. Phys.}\ }\textbf {\bibinfo {volume} {18}},\ \bibinfo {pages} {083044}
  (\bibinfo {year} {2016})}\BibitemShut {NoStop}%
\bibitem [{\citenamefont {Vieira}\ \emph {et~al.}(2013)\citenamefont {Vieira},
  \citenamefont {Amorim},\ and\ \citenamefont {Rigolin}}]{Vieira2013}%
  \BibitemOpen
  \bibfield  {author} {\bibinfo {author} {\bibfnamefont {R.}~\bibnamefont
  {Vieira}}, \bibinfo {author} {\bibfnamefont {E.~P.~M.}\ \bibnamefont
  {Amorim}}, \ and\ \bibinfo {author} {\bibfnamefont {G.}~\bibnamefont
  {Rigolin}},\ }\href {\doibase 10.1103/PhysRevLett.111.180503} {\bibfield
  {journal} {\bibinfo  {journal} {Phys. Rev. Lett.}\ }\textbf {\bibinfo
  {volume} {111}},\ \bibinfo {pages} {180503} (\bibinfo {year}
  {2013})}\BibitemShut {NoStop}%
\bibitem [{\citenamefont {Vieira}\ \emph {et~al.}(2014)\citenamefont {Vieira},
  \citenamefont {Amorim},\ and\ \citenamefont {Rigolin}}]{Vieira2014}%
  \BibitemOpen
  \bibfield  {author} {\bibinfo {author} {\bibfnamefont {R.}~\bibnamefont
  {Vieira}}, \bibinfo {author} {\bibfnamefont {E.~P.~M.}\ \bibnamefont
  {Amorim}}, \ and\ \bibinfo {author} {\bibfnamefont {G.}~\bibnamefont
  {Rigolin}},\ }\href {https://link.aps.org/doi/10.1103/PhysRevA.89.042307}
  {\bibfield  {journal} {\bibinfo  {journal} {Phys. Rev. A}\ }\textbf {\bibinfo
  {volume} {89}},\ \bibinfo {pages} {042307} (\bibinfo {year}
  {2014})}\BibitemShut {NoStop}%
\bibitem [{\citenamefont {Zeng}\ and\ \citenamefont {Yong}(2017)}]{Zeng2017}%
  \BibitemOpen
  \bibfield  {author} {\bibinfo {author} {\bibfnamefont {M.}~\bibnamefont
  {Zeng}}\ and\ \bibinfo {author} {\bibfnamefont {E.~H.}\ \bibnamefont
  {Yong}},\ }\href {\doibase 10.1038/s41598-017-12077-0} {\bibfield  {journal}
  {\bibinfo  {journal} {Sci. Rep.}\ }\textbf {\bibinfo {volume} {7}},\ \bibinfo
  {pages} {12024} (\bibinfo {year} {2017})}\BibitemShut {NoStop}%
\bibitem [{\citenamefont {Binosi}\ \emph {et~al.}(2007)\citenamefont {Binosi},
  \citenamefont {De~Chiara}, \citenamefont {Montangero},\ and\ \citenamefont
  {Recati}}]{Binosi2007}%
  \BibitemOpen
  \bibfield  {author} {\bibinfo {author} {\bibfnamefont {D.}~\bibnamefont
  {Binosi}}, \bibinfo {author} {\bibfnamefont {G.}~\bibnamefont {De~Chiara}},
  \bibinfo {author} {\bibfnamefont {S.}~\bibnamefont {Montangero}}, \ and\
  \bibinfo {author} {\bibfnamefont {A.}~\bibnamefont {Recati}},\ }\href
  {\doibase 10.1103/PhysRevB.76.140405} {\bibfield  {journal} {\bibinfo
  {journal} {Phys. Rev. B}\ }\textbf {\bibinfo {volume} {76}},\ \bibinfo
  {pages} {140405} (\bibinfo {year} {2007})}\BibitemShut {NoStop}%
\bibitem [{\citenamefont {Aharonov}\ \emph {et~al.}(1993)\citenamefont
  {Aharonov}, \citenamefont {Davidovich},\ and\ \citenamefont
  {Zagury}}]{Aharonov1993}%
  \BibitemOpen
  \bibfield  {author} {\bibinfo {author} {\bibfnamefont {Y.}~\bibnamefont
  {Aharonov}}, \bibinfo {author} {\bibfnamefont {L.}~\bibnamefont
  {Davidovich}}, \ and\ \bibinfo {author} {\bibfnamefont {N.}~\bibnamefont
  {Zagury}},\ }\href {\doibase 10.1103/PhysRevA.48.1687} {\bibfield  {journal}
  {\bibinfo  {journal} {Phys. Rev. A}\ }\textbf {\bibinfo {volume} {48}},\
  \bibinfo {pages} {1687} (\bibinfo {year} {1993})}\BibitemShut {NoStop}%
\bibitem [{\citenamefont {Pearson}(1905)}]{Karl1905}%
  \BibitemOpen
  \bibfield  {author} {\bibinfo {author} {\bibfnamefont {K.}~\bibnamefont
  {Pearson}},\ }\href {\doibase 10.1038/072294b0} {\bibfield  {journal}
  {\bibinfo  {journal} {Nature}\ }\textbf {\bibinfo {volume} {72}},\ \bibinfo
  {pages} {294} (\bibinfo {year} {1905})}\BibitemShut {NoStop}%
\bibitem [{\citenamefont {Childs}(2009)}]{Childs2009}%
  \BibitemOpen
  \bibfield  {author} {\bibinfo {author} {\bibfnamefont {A.~M.}\ \bibnamefont
  {Childs}},\ }\href {\doibase Artn 180501 10.1103/Physrevlett.102.180501}
  {\bibfield  {journal} {\bibinfo  {journal} {Phys. Rev. Lett.}\ }\textbf
  {\bibinfo {volume} {102}},\ \bibinfo {pages} {180501} (\bibinfo {year}
  {2009})}\BibitemShut {NoStop}%
\bibitem [{\citenamefont {Childs}\ \emph {et~al.}(2013)\citenamefont {Childs},
  \citenamefont {Gosset},\ and\ \citenamefont {Webb}}]{Childs2013}%
  \BibitemOpen
  \bibfield  {author} {\bibinfo {author} {\bibfnamefont {A.~M.}\ \bibnamefont
  {Childs}}, \bibinfo {author} {\bibfnamefont {D.}~\bibnamefont {Gosset}}, \
  and\ \bibinfo {author} {\bibfnamefont {Z.}~\bibnamefont {Webb}},\ }\href
  {\doibase 10.1126/science.1229957} {\bibfield  {journal} {\bibinfo  {journal}
  {Science}\ }\textbf {\bibinfo {volume} {339}},\ \bibinfo {pages} {791}
  (\bibinfo {year} {2013})}\BibitemShut {NoStop}%
\bibitem [{\citenamefont {Chandrashekar}(2013)}]{Chandrashekar2013}%
  \BibitemOpen
  \bibfield  {author} {\bibinfo {author} {\bibfnamefont {C.~M.}\ \bibnamefont
  {Chandrashekar}},\ }\href {\doibase 10.1038/srep02829} {\bibfield  {journal}
  {\bibinfo  {journal} {Sci. Rep.}\ }\textbf {\bibinfo {volume} {3}},\ \bibinfo
  {pages} {2829} (\bibinfo {year} {2013})}\BibitemShut {NoStop}%
\bibitem [{\citenamefont {Kitagawa}\ \emph {et~al.}(2010)\citenamefont
  {Kitagawa}, \citenamefont {Rudner}, \citenamefont {Berg},\ and\ \citenamefont
  {Demler}}]{Kitagawa2010}%
  \BibitemOpen
  \bibfield  {author} {\bibinfo {author} {\bibfnamefont {T.}~\bibnamefont
  {Kitagawa}}, \bibinfo {author} {\bibfnamefont {M.~S.}\ \bibnamefont
  {Rudner}}, \bibinfo {author} {\bibfnamefont {E.}~\bibnamefont {Berg}}, \ and\
  \bibinfo {author} {\bibfnamefont {E.}~\bibnamefont {Demler}},\ }\href
  {\doibase 10.1103/PhysRevA.82.033429} {\bibfield  {journal} {\bibinfo
  {journal} {Phys. Rev. A}\ }\textbf {\bibinfo {volume} {82}},\ \bibinfo
  {pages} {033429} (\bibinfo {year} {2010})}\BibitemShut {NoStop}%
\bibitem [{\citenamefont {Kitagawa}\ \emph {et~al.}(2012)\citenamefont
  {Kitagawa}, \citenamefont {Broome}, \citenamefont {Fedrizzi}, \citenamefont
  {Rudner}, \citenamefont {Berg}, \citenamefont {Kassal}, \citenamefont
  {Aspuru-Guzik}, \citenamefont {Demler},\ and\ \citenamefont
  {White}}]{Kitagawa2012}%
  \BibitemOpen
  \bibfield  {author} {\bibinfo {author} {\bibfnamefont {T.}~\bibnamefont
  {Kitagawa}}, \bibinfo {author} {\bibfnamefont {M.~A.}\ \bibnamefont
  {Broome}}, \bibinfo {author} {\bibfnamefont {A.}~\bibnamefont {Fedrizzi}},
  \bibinfo {author} {\bibfnamefont {M.~S.}\ \bibnamefont {Rudner}}, \bibinfo
  {author} {\bibfnamefont {E.}~\bibnamefont {Berg}}, \bibinfo {author}
  {\bibfnamefont {I.}~\bibnamefont {Kassal}}, \bibinfo {author} {\bibfnamefont
  {A.}~\bibnamefont {Aspuru-Guzik}}, \bibinfo {author} {\bibfnamefont
  {E.}~\bibnamefont {Demler}}, \ and\ \bibinfo {author} {\bibfnamefont {A.~G.}\
  \bibnamefont {White}},\ }\href {\doibase 10.1038/Ncomms1872} {\bibfield
  {journal} {\bibinfo  {journal} {Nat. Commun.}\ }\textbf {\bibinfo {volume}
  {3}},\ \bibinfo {pages} {882} (\bibinfo {year} {2012})}\BibitemShut {NoStop}%
\bibitem [{\citenamefont {Rechtsman}\ \emph {et~al.}(2016)\citenamefont
  {Rechtsman}, \citenamefont {Lumer}, \citenamefont {Plotnik}, \citenamefont
  {Perez-Leija}, \citenamefont {Szameit},\ and\ \citenamefont
  {Segev}}]{rechtsman2016}%
  \BibitemOpen
  \bibfield  {author} {\bibinfo {author} {\bibfnamefont {M.~C.}\ \bibnamefont
  {Rechtsman}}, \bibinfo {author} {\bibfnamefont {Y.}~\bibnamefont {Lumer}},
  \bibinfo {author} {\bibfnamefont {Y.}~\bibnamefont {Plotnik}}, \bibinfo
  {author} {\bibfnamefont {A.}~\bibnamefont {Perez-Leija}}, \bibinfo {author}
  {\bibfnamefont {A.}~\bibnamefont {Szameit}}, \ and\ \bibinfo {author}
  {\bibfnamefont {M.}~\bibnamefont {Segev}},\ }\href {\doibase
  10.1364/OPTICA.3.000925} {\bibfield  {journal} {\bibinfo  {journal} {Optica}\
  }\textbf {\bibinfo {volume} {3}},\ \bibinfo {pages} {925} (\bibinfo {year}
  {2016})}\BibitemShut {NoStop}%
\bibitem [{\citenamefont {Blanchard}\ and\ \citenamefont
  {Hongler}(2004)}]{Blanchard2004}%
  \BibitemOpen
  \bibfield  {author} {\bibinfo {author} {\bibfnamefont {P.}~\bibnamefont
  {Blanchard}}\ and\ \bibinfo {author} {\bibfnamefont {M.-O.}\ \bibnamefont
  {Hongler}},\ }\href {\doibase 10.1103/PhysRevLett.92.120601} {\bibfield
  {journal} {\bibinfo  {journal} {Phys. Rev. Lett.}\ }\textbf {\bibinfo
  {volume} {92}},\ \bibinfo {pages} {120601} (\bibinfo {year}
  {2004})}\BibitemShut {NoStop}%
\bibitem [{\citenamefont {Mackay}\ \emph {et~al.}(2002)\citenamefont {Mackay},
  \citenamefont {Bartlett}, \citenamefont {Stephenson},\ and\ \citenamefont
  {Sanders}}]{Mackay2002}%
  \BibitemOpen
  \bibfield  {author} {\bibinfo {author} {\bibfnamefont {T.~D.}\ \bibnamefont
  {Mackay}}, \bibinfo {author} {\bibfnamefont {S.~D.}\ \bibnamefont
  {Bartlett}}, \bibinfo {author} {\bibfnamefont {L.~T.}\ \bibnamefont
  {Stephenson}}, \ and\ \bibinfo {author} {\bibfnamefont {B.~C.}\ \bibnamefont
  {Sanders}},\ }\href {http://stacks.iop.org/0305-4470/35/i=12/a=304}
  {\bibfield  {journal} {\bibinfo  {journal} {J. Phys. A-Math. Gen.}\ }\textbf
  {\bibinfo {volume} {35}},\ \bibinfo {pages} {2745} (\bibinfo {year}
  {2002})}\BibitemShut {NoStop}%
\bibitem [{\citenamefont {Carneiro}\ \emph {et~al.}(2005)\citenamefont
  {Carneiro}, \citenamefont {Loo}, \citenamefont {Xu}, \citenamefont {Girerd},
  \citenamefont {Kendon},\ and\ \citenamefont {Knight}}]{Carneiro2005}%
  \BibitemOpen
  \bibfield  {author} {\bibinfo {author} {\bibfnamefont {I.}~\bibnamefont
  {Carneiro}}, \bibinfo {author} {\bibfnamefont {M.}~\bibnamefont {Loo}},
  \bibinfo {author} {\bibfnamefont {X.}~\bibnamefont {Xu}}, \bibinfo {author}
  {\bibfnamefont {M.}~\bibnamefont {Girerd}}, \bibinfo {author} {\bibfnamefont
  {V.}~\bibnamefont {Kendon}}, \ and\ \bibinfo {author} {\bibfnamefont {P.~L.}\
  \bibnamefont {Knight}},\ }\href {\doibase 10.1088/1367-2630/7/1/156}
  {\bibfield  {journal} {\bibinfo  {journal} {New J. Phys.}\ }\textbf {\bibinfo
  {volume} {7}},\ \bibinfo {pages} {156} (\bibinfo {year} {2005})}\BibitemShut
  {NoStop}%
\bibitem [{\citenamefont {Di~Franco}\ \emph {et~al.}(2011)\citenamefont
  {Di~Franco}, \citenamefont {Mc~Gettrick},\ and\ \citenamefont
  {Busch}}]{Franco2011}%
  \BibitemOpen
  \bibfield  {author} {\bibinfo {author} {\bibfnamefont {C.}~\bibnamefont
  {Di~Franco}}, \bibinfo {author} {\bibfnamefont {M.}~\bibnamefont
  {Mc~Gettrick}}, \ and\ \bibinfo {author} {\bibfnamefont {T.}~\bibnamefont
  {Busch}},\ }\href {https://link.aps.org/doi/10.1103/PhysRevLett.106.080502}
  {\bibfield  {journal} {\bibinfo  {journal} {Phys. Rev. Lett.}\ }\textbf
  {\bibinfo {volume} {106}},\ \bibinfo {pages} {080502} (\bibinfo {year}
  {2011})}\BibitemShut {NoStop}%
\bibitem [{\citenamefont {Venegas-Andraca}(2012)}]{Venegas2012}%
  \BibitemOpen
  \bibfield  {author} {\bibinfo {author} {\bibfnamefont {S.~E.}\ \bibnamefont
  {Venegas-Andraca}},\ }\href {\doibase 10.1007/s11128-012-0432-5} {\bibfield
  {journal} {\bibinfo  {journal} {Quantum Inf. Processing}\ }\textbf {\bibinfo
  {volume} {11}},\ \bibinfo {pages} {1015} (\bibinfo {year}
  {2012})}\BibitemShut {NoStop}%
\bibitem [{\citenamefont {Morin}\ \emph {et~al.}(2013)\citenamefont {Morin},
  \citenamefont {Bancal}, \citenamefont {Ho}, \citenamefont {Sekatski},
  \citenamefont {D’Auria}, \citenamefont {Gisin}, \citenamefont {Laurat},\
  and\ \citenamefont {Sangouard}}]{Morin2013}%
  \BibitemOpen
  \bibfield  {author} {\bibinfo {author} {\bibfnamefont {O.}~\bibnamefont
  {Morin}}, \bibinfo {author} {\bibfnamefont {J.~D.}\ \bibnamefont {Bancal}},
  \bibinfo {author} {\bibfnamefont {M.}~\bibnamefont {Ho}}, \bibinfo {author}
  {\bibfnamefont {P.}~\bibnamefont {Sekatski}}, \bibinfo {author}
  {\bibfnamefont {V.}~\bibnamefont {D’Auria}}, \bibinfo {author}
  {\bibfnamefont {N.}~\bibnamefont {Gisin}}, \bibinfo {author} {\bibfnamefont
  {J.}~\bibnamefont {Laurat}}, \ and\ \bibinfo {author} {\bibfnamefont
  {N.}~\bibnamefont {Sangouard}},\ }\href {\doibase
  10.1103/PhysRevLett.110.130401} {\bibfield  {journal} {\bibinfo  {journal}
  {Phys. Rev. Lett.}\ }\textbf {\bibinfo {volume} {110}},\ \bibinfo {pages}
  {130401} (\bibinfo {year} {2013})}\BibitemShut {NoStop}%
\bibitem [{\citenamefont {Choi}\ \emph {et~al.}(2008)\citenamefont {Choi},
  \citenamefont {Deng}, \citenamefont {Laurat},\ and\ \citenamefont
  {Kimble}}]{Choi2008}%
  \BibitemOpen
  \bibfield  {author} {\bibinfo {author} {\bibfnamefont {K.~S.}\ \bibnamefont
  {Choi}}, \bibinfo {author} {\bibfnamefont {H.}~\bibnamefont {Deng}}, \bibinfo
  {author} {\bibfnamefont {J.}~\bibnamefont {Laurat}}, \ and\ \bibinfo {author}
  {\bibfnamefont {H.~J.}\ \bibnamefont {Kimble}},\ }\href {\doibase
  10.1038/nature06670} {\bibfield  {journal} {\bibinfo  {journal} {Nature}\
  }\textbf {\bibinfo {volume} {452}},\ \bibinfo {pages} {67} (\bibinfo {year}
  {2008})}\BibitemShut {NoStop}%
\bibitem [{\citenamefont {Cardano}\ \emph {et~al.}(2015)\citenamefont
  {Cardano}, \citenamefont {Massa}, \citenamefont {Qassim}, \citenamefont
  {Karimi}, \citenamefont {Slussarenko}, \citenamefont {Paparo}, \citenamefont
  {de~Lisio}, \citenamefont {Sciarrino}, \citenamefont {Santamato},
  \citenamefont {Boyd},\ and\ \citenamefont {Marrucci}}]{Cardano2015}%
  \BibitemOpen
  \bibfield  {author} {\bibinfo {author} {\bibfnamefont {F.}~\bibnamefont
  {Cardano}}, \bibinfo {author} {\bibfnamefont {F.}~\bibnamefont {Massa}},
  \bibinfo {author} {\bibfnamefont {H.}~\bibnamefont {Qassim}}, \bibinfo
  {author} {\bibfnamefont {E.}~\bibnamefont {Karimi}}, \bibinfo {author}
  {\bibfnamefont {S.}~\bibnamefont {Slussarenko}}, \bibinfo {author}
  {\bibfnamefont {D.}~\bibnamefont {Paparo}}, \bibinfo {author} {\bibfnamefont
  {C.}~\bibnamefont {de~Lisio}}, \bibinfo {author} {\bibfnamefont
  {F.}~\bibnamefont {Sciarrino}}, \bibinfo {author} {\bibfnamefont
  {E.}~\bibnamefont {Santamato}}, \bibinfo {author} {\bibfnamefont {R.~W.}\
  \bibnamefont {Boyd}}, \ and\ \bibinfo {author} {\bibfnamefont
  {L.}~\bibnamefont {Marrucci}},\ }\href {\doibase 10.1126/sciadv.1500087}
  {\bibfield  {journal} {\bibinfo  {journal} {Sci. Adv.}\ }\textbf {\bibinfo
  {volume} {1}},\ \bibinfo {pages} {1500087} (\bibinfo {year}
  {2015})}\BibitemShut {NoStop}%
\bibitem [{\citenamefont {Schreiber}\ \emph {et~al.}(2010)\citenamefont
  {Schreiber}, \citenamefont {Cassemiro}, \citenamefont {Potocek},
  \citenamefont {Gabris}, \citenamefont {Mosley}, \citenamefont {Andersson},
  \citenamefont {Jex},\ and\ \citenamefont {Silberhorn}}]{Schreiber2010}%
  \BibitemOpen
  \bibfield  {author} {\bibinfo {author} {\bibfnamefont {A.}~\bibnamefont
  {Schreiber}}, \bibinfo {author} {\bibfnamefont {K.~N.}\ \bibnamefont
  {Cassemiro}}, \bibinfo {author} {\bibfnamefont {V.}~\bibnamefont {Potocek}},
  \bibinfo {author} {\bibfnamefont {A.}~\bibnamefont {Gabris}}, \bibinfo
  {author} {\bibfnamefont {P.~J.}\ \bibnamefont {Mosley}}, \bibinfo {author}
  {\bibfnamefont {E.}~\bibnamefont {Andersson}}, \bibinfo {author}
  {\bibfnamefont {I.}~\bibnamefont {Jex}}, \ and\ \bibinfo {author}
  {\bibfnamefont {C.}~\bibnamefont {Silberhorn}},\ }\href {\doibase Artn 050502
  10.1103/Physrevlett.104.050502} {\bibfield  {journal} {\bibinfo  {journal}
  {Phys. Rev. Lett.}\ }\textbf {\bibinfo {volume} {104}},\ \bibinfo {pages}
  {050502} (\bibinfo {year} {2010})}\BibitemShut {NoStop}%
\bibitem [{\citenamefont {Schreiber}\ \emph {et~al.}(2011)\citenamefont
  {Schreiber}, \citenamefont {Cassemiro}, \citenamefont {Potoček},
  \citenamefont {Gabris}, \citenamefont {Jex},\ and\ \citenamefont
  {Silberhorn}}]{Schreiber2011}%
  \BibitemOpen
  \bibfield  {author} {\bibinfo {author} {\bibfnamefont {A.}~\bibnamefont
  {Schreiber}}, \bibinfo {author} {\bibfnamefont {K.~N.}\ \bibnamefont
  {Cassemiro}}, \bibinfo {author} {\bibfnamefont {V.}~\bibnamefont {Potoček}},
  \bibinfo {author} {\bibfnamefont {A.}~\bibnamefont {Gabris}}, \bibinfo
  {author} {\bibfnamefont {I.}~\bibnamefont {Jex}}, \ and\ \bibinfo {author}
  {\bibfnamefont {C.}~\bibnamefont {Silberhorn}},\ }\href
  {https://link.aps.org/doi/10.1103/PhysRevLett.106.180403} {\bibfield
  {journal} {\bibinfo  {journal} {Phys. Rev. Lett.}\ }\textbf {\bibinfo
  {volume} {106}},\ \bibinfo {pages} {180403} (\bibinfo {year}
  {2011})}\BibitemShut {NoStop}%
\bibitem [{\citenamefont {Sansoni}\ \emph {et~al.}(2012)\citenamefont
  {Sansoni}, \citenamefont {Sciarrino}, \citenamefont {Vallone}, \citenamefont
  {Mataloni}, \citenamefont {Crespi}, \citenamefont {Ramponi},\ and\
  \citenamefont {Osellame}}]{Sansoni2012}%
  \BibitemOpen
  \bibfield  {author} {\bibinfo {author} {\bibfnamefont {L.}~\bibnamefont
  {Sansoni}}, \bibinfo {author} {\bibfnamefont {F.}~\bibnamefont {Sciarrino}},
  \bibinfo {author} {\bibfnamefont {G.}~\bibnamefont {Vallone}}, \bibinfo
  {author} {\bibfnamefont {P.}~\bibnamefont {Mataloni}}, \bibinfo {author}
  {\bibfnamefont {A.}~\bibnamefont {Crespi}}, \bibinfo {author} {\bibfnamefont
  {R.}~\bibnamefont {Ramponi}}, \ and\ \bibinfo {author} {\bibfnamefont
  {R.}~\bibnamefont {Osellame}},\ }\href {\doibase
  10.1103/PhysRevLett.108.010502} {\bibfield  {journal} {\bibinfo  {journal}
  {Phys. Rev. Lett.}\ }\textbf {\bibinfo {volume} {108}},\ \bibinfo {pages}
  {010502} (\bibinfo {year} {2012})}\BibitemShut {NoStop}%
\bibitem [{\citenamefont {Peruzzo}\ \emph {et~al.}(2010)\citenamefont
  {Peruzzo}, \citenamefont {Lobino}, \citenamefont {Matthews}, \citenamefont
  {Matsuda}, \citenamefont {Politi}, \citenamefont {Poulios}, \citenamefont
  {Zhou}, \citenamefont {Lahini}, \citenamefont {Ismail}, \citenamefont
  {Worhoff}, \citenamefont {Bromberg}, \citenamefont {Silberberg},
  \citenamefont {Thompson},\ and\ \citenamefont {O'Brien}}]{Peruzzo2010}%
  \BibitemOpen
  \bibfield  {author} {\bibinfo {author} {\bibfnamefont {A.}~\bibnamefont
  {Peruzzo}}, \bibinfo {author} {\bibfnamefont {M.}~\bibnamefont {Lobino}},
  \bibinfo {author} {\bibfnamefont {J.~C.~F.}\ \bibnamefont {Matthews}},
  \bibinfo {author} {\bibfnamefont {N.}~\bibnamefont {Matsuda}}, \bibinfo
  {author} {\bibfnamefont {A.}~\bibnamefont {Politi}}, \bibinfo {author}
  {\bibfnamefont {K.}~\bibnamefont {Poulios}}, \bibinfo {author} {\bibfnamefont
  {X.~Q.}\ \bibnamefont {Zhou}}, \bibinfo {author} {\bibfnamefont
  {Y.}~\bibnamefont {Lahini}}, \bibinfo {author} {\bibfnamefont
  {N.}~\bibnamefont {Ismail}}, \bibinfo {author} {\bibfnamefont
  {K.}~\bibnamefont {Worhoff}}, \bibinfo {author} {\bibfnamefont
  {Y.}~\bibnamefont {Bromberg}}, \bibinfo {author} {\bibfnamefont
  {Y.}~\bibnamefont {Silberberg}}, \bibinfo {author} {\bibfnamefont {M.~G.}\
  \bibnamefont {Thompson}}, \ and\ \bibinfo {author} {\bibfnamefont {J.~L.}\
  \bibnamefont {O'Brien}},\ }\href {\doibase 10.1126/science.1193515}
  {\bibfield  {journal} {\bibinfo  {journal} {Science}\ }\textbf {\bibinfo
  {volume} {329}},\ \bibinfo {pages} {1500} (\bibinfo {year}
  {2010})}\BibitemShut {NoStop}%
\bibitem [{\citenamefont {Titchener}\ \emph {et~al.}(2016)\citenamefont
  {Titchener}, \citenamefont {Solntsev},\ and\ \citenamefont
  {Sukhorukov}}]{titchener2016}%
  \BibitemOpen
  \bibfield  {author} {\bibinfo {author} {\bibfnamefont {J.~G.}\ \bibnamefont
  {Titchener}}, \bibinfo {author} {\bibfnamefont {A.~S.}\ \bibnamefont
  {Solntsev}}, \ and\ \bibinfo {author} {\bibfnamefont {A.~A.}\ \bibnamefont
  {Sukhorukov}},\ }\href {\doibase 10.1364/OL.41.004079} {\bibfield  {journal}
  {\bibinfo  {journal} {Opt. Lett.}\ }\textbf {\bibinfo {volume} {41}},\
  \bibinfo {pages} {4079} (\bibinfo {year} {2016})}\BibitemShut {NoStop}%
\bibitem [{\citenamefont {Perets}\ \emph {et~al.}(2008)\citenamefont {Perets},
  \citenamefont {Lahini}, \citenamefont {Pozzi}, \citenamefont {Sorel},
  \citenamefont {Morandotti},\ and\ \citenamefont {Silberberg}}]{Perets2008}%
  \BibitemOpen
  \bibfield  {author} {\bibinfo {author} {\bibfnamefont {H.~B.}\ \bibnamefont
  {Perets}}, \bibinfo {author} {\bibfnamefont {Y.}~\bibnamefont {Lahini}},
  \bibinfo {author} {\bibfnamefont {F.}~\bibnamefont {Pozzi}}, \bibinfo
  {author} {\bibfnamefont {M.}~\bibnamefont {Sorel}}, \bibinfo {author}
  {\bibfnamefont {R.}~\bibnamefont {Morandotti}}, \ and\ \bibinfo {author}
  {\bibfnamefont {Y.}~\bibnamefont {Silberberg}},\ }\href {\doibase
  10.1103/PhysRevLett.100.170506} {\bibfield  {journal} {\bibinfo  {journal}
  {Phys. Rev. Lett.}\ }\textbf {\bibinfo {volume} {100}},\ \bibinfo {pages}
  {170506} (\bibinfo {year} {2008})}\BibitemShut {NoStop}%
\bibitem [{\citenamefont {Regensburger}\ \emph {et~al.}(2011)\citenamefont
  {Regensburger}, \citenamefont {Bersch}, \citenamefont {Hinrichs},
  \citenamefont {Onishchukov}, \citenamefont {Schreiber}, \citenamefont
  {Silberhorn},\ and\ \citenamefont {Peschel}}]{Regensburger2011}%
  \BibitemOpen
  \bibfield  {author} {\bibinfo {author} {\bibfnamefont {A.}~\bibnamefont
  {Regensburger}}, \bibinfo {author} {\bibfnamefont {C.}~\bibnamefont
  {Bersch}}, \bibinfo {author} {\bibfnamefont {B.}~\bibnamefont {Hinrichs}},
  \bibinfo {author} {\bibfnamefont {G.}~\bibnamefont {Onishchukov}}, \bibinfo
  {author} {\bibfnamefont {A.}~\bibnamefont {Schreiber}}, \bibinfo {author}
  {\bibfnamefont {C.}~\bibnamefont {Silberhorn}}, \ and\ \bibinfo {author}
  {\bibfnamefont {U.}~\bibnamefont {Peschel}},\ }\href {\doibase
  10.1103/PhysRevLett.107.233902} {\bibfield  {journal} {\bibinfo  {journal}
  {Phys. Rev. Lett.}\ }\textbf {\bibinfo {volume} {107}},\ \bibinfo {pages}
  {233902} (\bibinfo {year} {2011})}\BibitemShut {NoStop}%
\bibitem [{\citenamefont {Schreiber}\ \emph {et~al.}(2012)\citenamefont
  {Schreiber}, \citenamefont {Gábris}, \citenamefont {Rohde}, \citenamefont
  {Laiho}, \citenamefont {Štefaňák}, \citenamefont {Potoček}, \citenamefont
  {Hamilton}, \citenamefont {Jex},\ and\ \citenamefont
  {Silberhorn}}]{Schreiber2012}%
  \BibitemOpen
  \bibfield  {author} {\bibinfo {author} {\bibfnamefont {A.}~\bibnamefont
  {Schreiber}}, \bibinfo {author} {\bibfnamefont {A.}~\bibnamefont {Gábris}},
  \bibinfo {author} {\bibfnamefont {P.~P.}\ \bibnamefont {Rohde}}, \bibinfo
  {author} {\bibfnamefont {K.}~\bibnamefont {Laiho}}, \bibinfo {author}
  {\bibfnamefont {M.}~\bibnamefont {Štefaňák}}, \bibinfo {author}
  {\bibfnamefont {V.}~\bibnamefont {Potoček}}, \bibinfo {author}
  {\bibfnamefont {C.}~\bibnamefont {Hamilton}}, \bibinfo {author}
  {\bibfnamefont {I.}~\bibnamefont {Jex}}, \ and\ \bibinfo {author}
  {\bibfnamefont {C.}~\bibnamefont {Silberhorn}},\ }\href {\doibase
  10.1126/science.1218448} {\bibfield  {journal} {\bibinfo  {journal}
  {Science}\ }\textbf {\bibinfo {volume} {336}},\ \bibinfo {pages} {55}
  (\bibinfo {year} {2012})}\BibitemShut {NoStop}%
\bibitem [{\citenamefont {Boutari}\ \emph {et~al.}(2016)\citenamefont
  {Boutari}, \citenamefont {Feizpour}, \citenamefont {Barz}, \citenamefont
  {Di~Franco}, \citenamefont {Kim}, \citenamefont {Kolthammer},\ and\
  \citenamefont {Walmsley}}]{Boutari2016}%
  \BibitemOpen
  \bibfield  {author} {\bibinfo {author} {\bibfnamefont {J.}~\bibnamefont
  {Boutari}}, \bibinfo {author} {\bibfnamefont {A.}~\bibnamefont {Feizpour}},
  \bibinfo {author} {\bibfnamefont {S.}~\bibnamefont {Barz}}, \bibinfo {author}
  {\bibfnamefont {C.}~\bibnamefont {Di~Franco}}, \bibinfo {author}
  {\bibfnamefont {M.~S.}\ \bibnamefont {Kim}}, \bibinfo {author} {\bibfnamefont
  {W.~S.}\ \bibnamefont {Kolthammer}}, \ and\ \bibinfo {author} {\bibfnamefont
  {I.~A.}\ \bibnamefont {Walmsley}},\ }\href {\doibase
  10.1088/2040-8978/18/9/094007} {\bibfield  {journal} {\bibinfo  {journal} {J.
  Opt.}\ }\textbf {\bibinfo {volume} {18}},\ \bibinfo {pages} {094007}
  (\bibinfo {year} {2016})}\BibitemShut {NoStop}%
\bibitem [{\citenamefont {Karski}\ \emph {et~al.}(2009)\citenamefont {Karski},
  \citenamefont {Förster}, \citenamefont {Choi}, \citenamefont {Steffen},
  \citenamefont {Alt}, \citenamefont {Meschede},\ and\ \citenamefont
  {Widera}}]{Karski2009}%
  \BibitemOpen
  \bibfield  {author} {\bibinfo {author} {\bibfnamefont {M.}~\bibnamefont
  {Karski}}, \bibinfo {author} {\bibfnamefont {L.}~\bibnamefont {Förster}},
  \bibinfo {author} {\bibfnamefont {J.-M.}\ \bibnamefont {Choi}}, \bibinfo
  {author} {\bibfnamefont {A.}~\bibnamefont {Steffen}}, \bibinfo {author}
  {\bibfnamefont {W.}~\bibnamefont {Alt}}, \bibinfo {author} {\bibfnamefont
  {D.}~\bibnamefont {Meschede}}, \ and\ \bibinfo {author} {\bibfnamefont
  {A.}~\bibnamefont {Widera}},\ }\href {\doibase 10.1126/science.1174436}
  {\bibfield  {journal} {\bibinfo  {journal} {Science}\ }\textbf {\bibinfo
  {volume} {325}},\ \bibinfo {pages} {174} (\bibinfo {year}
  {2009})}\BibitemShut {NoStop}%
\bibitem [{\citenamefont {Xu}\ \emph {et~al.}(2018)\citenamefont {Xu},
  \citenamefont {Wang}, \citenamefont {Pan}, \citenamefont {Sun}, \citenamefont
  {Xu}, \citenamefont {Chen}, \citenamefont {Tang}, \citenamefont {Gong},
  \citenamefont {Han}, \citenamefont {Li},\ and\ \citenamefont {Guo}}]{Xu2018}%
  \BibitemOpen
  \bibfield  {author} {\bibinfo {author} {\bibfnamefont {X.-Y.}\ \bibnamefont
  {Xu}}, \bibinfo {author} {\bibfnamefont {Q.-Q.}\ \bibnamefont {Wang}},
  \bibinfo {author} {\bibfnamefont {W.-W.}\ \bibnamefont {Pan}}, \bibinfo
  {author} {\bibfnamefont {K.}~\bibnamefont {Sun}}, \bibinfo {author}
  {\bibfnamefont {J.-S.}\ \bibnamefont {Xu}}, \bibinfo {author} {\bibfnamefont
  {G.}~\bibnamefont {Chen}}, \bibinfo {author} {\bibfnamefont {J.-S.}\
  \bibnamefont {Tang}}, \bibinfo {author} {\bibfnamefont {M.}~\bibnamefont
  {Gong}}, \bibinfo {author} {\bibfnamefont {Y.-J.}\ \bibnamefont {Han}},
  \bibinfo {author} {\bibfnamefont {C.-F.}\ \bibnamefont {Li}}, \ and\ \bibinfo
  {author} {\bibfnamefont {G.-C.}\ \bibnamefont {Guo}},\ }\href {\doibase
  10.1103/PhysRevLett.120.260501} {\bibfield  {journal} {\bibinfo  {journal}
  {Phys. Rev. Lett.}\ }\textbf {\bibinfo {volume} {120}},\ \bibinfo {pages}
  {260501} (\bibinfo {year} {2018})}\BibitemShut {NoStop}%
\bibitem [{\citenamefont {Barkhofen}\ \emph {et~al.}(2017)\citenamefont
  {Barkhofen}, \citenamefont {Nitsche}, \citenamefont {Elster}, \citenamefont
  {Lorz}, \citenamefont {Gábris}, \citenamefont {Jex},\ and\ \citenamefont
  {Silberhorn}}]{Barkhofen2017}%
  \BibitemOpen
  \bibfield  {author} {\bibinfo {author} {\bibfnamefont {S.}~\bibnamefont
  {Barkhofen}}, \bibinfo {author} {\bibfnamefont {T.}~\bibnamefont {Nitsche}},
  \bibinfo {author} {\bibfnamefont {F.}~\bibnamefont {Elster}}, \bibinfo
  {author} {\bibfnamefont {L.}~\bibnamefont {Lorz}}, \bibinfo {author}
  {\bibfnamefont {A.}~\bibnamefont {Gábris}}, \bibinfo {author} {\bibfnamefont
  {I.}~\bibnamefont {Jex}}, \ and\ \bibinfo {author} {\bibfnamefont
  {C.}~\bibnamefont {Silberhorn}},\ }\href
  {https://link.aps.org/doi/10.1103/PhysRevA.96.033846} {\bibfield  {journal}
  {\bibinfo  {journal} {Phys. Rev. A}\ }\textbf {\bibinfo {volume} {96}},\
  \bibinfo {pages} {033846} (\bibinfo {year} {2017})}\BibitemShut {NoStop}%
\bibitem [{\citenamefont {Eichelkraut}\ \emph {et~al.}(2013)\citenamefont
  {Eichelkraut}, \citenamefont {Heilmann}, \citenamefont {Weimann},
  \citenamefont {St{\"u}tzer}, \citenamefont {Dreisow}, \citenamefont
  {Christodoulides}, \citenamefont {Nolte},\ and\ \citenamefont
  {Szameit}}]{Eichelkraut2013}%
  \BibitemOpen
  \bibfield  {author} {\bibinfo {author} {\bibfnamefont {T.}~\bibnamefont
  {Eichelkraut}}, \bibinfo {author} {\bibfnamefont {R.}~\bibnamefont
  {Heilmann}}, \bibinfo {author} {\bibfnamefont {S.}~\bibnamefont {Weimann}},
  \bibinfo {author} {\bibfnamefont {S.}~\bibnamefont {St{\"u}tzer}}, \bibinfo
  {author} {\bibfnamefont {F.}~\bibnamefont {Dreisow}}, \bibinfo {author}
  {\bibfnamefont {D.~N.}\ \bibnamefont {Christodoulides}}, \bibinfo {author}
  {\bibfnamefont {S.}~\bibnamefont {Nolte}}, \ and\ \bibinfo {author}
  {\bibfnamefont {A.}~\bibnamefont {Szameit}},\ }\href {\doibase
  10.1038/ncomms3533} {\bibfield  {journal} {\bibinfo  {journal} {Nat.
  Commun.}\ }\textbf {\bibinfo {volume} {4}},\ \bibinfo {pages} {2533}
  (\bibinfo {year} {2013})}\BibitemShut {NoStop}%
\bibitem [{\citenamefont {Ribeiro}\ \emph {et~al.}(2004)\citenamefont
  {Ribeiro}, \citenamefont {Milman},\ and\ \citenamefont
  {Mosseri}}]{Ribeiro2004}%
  \BibitemOpen
  \bibfield  {author} {\bibinfo {author} {\bibfnamefont {P.}~\bibnamefont
  {Ribeiro}}, \bibinfo {author} {\bibfnamefont {P.}~\bibnamefont {Milman}}, \
  and\ \bibinfo {author} {\bibfnamefont {R.}~\bibnamefont {Mosseri}},\ }\href
  {\doibase 10.1103/PhysRevLett.93.190503} {\bibfield  {journal} {\bibinfo
  {journal} {Phys. Rev. Lett.}\ }\textbf {\bibinfo {volume} {93}},\ \bibinfo
  {pages} {190503} (\bibinfo {year} {2004})}\BibitemShut {NoStop}%
\bibitem [{\citenamefont {Lempel}\ and\ \citenamefont
  {Ziv}(1976)}]{Lempel1976}%
  \BibitemOpen
  \bibfield  {author} {\bibinfo {author} {\bibfnamefont {A.}~\bibnamefont
  {Lempel}}\ and\ \bibinfo {author} {\bibfnamefont {J.}~\bibnamefont {Ziv}},\
  }\href {\doibase 10.1109/TIT.1976.1055501} {\bibfield  {journal} {\bibinfo
  {journal} {IEEE Trans. On IT}\ }\textbf {\bibinfo {volume} {22}},\ \bibinfo
  {pages} {75} (\bibinfo {year} {1976})}\BibitemShut {NoStop}%
\bibitem [{\citenamefont {Crutchfield}(2011)}]{Crutchfield2011}%
  \BibitemOpen
  \bibfield  {author} {\bibinfo {author} {\bibfnamefont {J.~P.}\ \bibnamefont
  {Crutchfield}},\ }\href {\doibase 10.1038/nphys2190} {\bibfield  {journal}
  {\bibinfo  {journal} {Nat. Phys.}\ }\textbf {\bibinfo {volume} {8}},\
  \bibinfo {pages} {17} (\bibinfo {year} {2011})}\BibitemShut {NoStop}%
\bibitem [{\citenamefont {Hu}\ \emph {et~al.}(2006)\citenamefont {Hu},
  \citenamefont {Gao},\ and\ \citenamefont {Principe}}]{hu2006}%
  \BibitemOpen
  \bibfield  {author} {\bibinfo {author} {\bibfnamefont {J.}~\bibnamefont
  {Hu}}, \bibinfo {author} {\bibfnamefont {J.}~\bibnamefont {Gao}}, \ and\
  \bibinfo {author} {\bibfnamefont {J.~C.}\ \bibnamefont {Principe}},\ }\href
  {\doibase 10.1109/TBME.2006.883825} {\bibfield  {journal} {\bibinfo
  {journal} {IEEE Trans. Biomed. Eng.}\ }\textbf {\bibinfo {volume} {53}},\
  \bibinfo {pages} {2606} (\bibinfo {year} {2006})}\BibitemShut {NoStop}%
\end{thebibliography}%

\newpage
\appendix
\section{Supplementary for \emph{Dynamic-Disorder-Induced Enhancement of Entanglement in Photonic Quantum Walks}}

\section{The details of the experiment setup}

\emph{Experimental time multiplexing photonic DTQWs ---}
In this work, the QWs is based on the time multiplexing protocol. However, for overcoming the problem of the extra loss, birefringence crystal are used to replace the asymmetric Mach-Zehnder interferometer and implement the spin-orbit coupling. Heralded single photons generated in SPDC are employed as the walker, the polarization is employed as the coin space, such that its polarization state can be rotated to any of the single qubit states by wave plates, and the arriving time of the photons, encoded in time bin, acts as the position space. One step quantum walk is realized by a module composed of a half-wave plate (HWP) or a quarter-wave plate (QWP) and one piece of birefringence crystal. The coin rotation operator implemented by the former is
\begin{equation}
\hat R_\text{HWP}(\theta) = e^{-i2\theta\hat\sigma_y}\hat\sigma_z,\hat R_\text{QWP}(\theta) = e^{-i\theta\hat\sigma_y}e^{-i(\pi/4)\hat\sigma_z}e^{i\theta\hat\sigma_y}.
\end{equation}
where $\theta$ is the rotation angle of the optical axis of the wave plate, and $\hat\sigma_i$ denotes the Pauli matrices such that the eigenstates of the coin are $|H\rangle$ and $|V\rangle$, corresponding to the horizontal and vertical polarizations respectively, with the condition $\hat\sigma_z|H\rangle = |H\rangle$ and $\hat\sigma_z|V\rangle = -|V\rangle$. The birefringence of the latter causes the horizontal components to travel faster inside the crystal than the vertical one. Then, after passing through the crystal, the photons in the horizontal polarization move a step forward. Considering the dispersion after passing through a large number of crystals and the fact that the time bin encoding the position of the walker in reality is a single pulse with a typical duration of a few hundred femtoseconds, such a shift in time should be sufficiently large to distinguish the neighborhood pulses at last. The magnitude of the polarization-dependent time shift by the birefringence crystal depends on the crystal length and the cut angle. In this experiment, for introducing as weak dispersion as possible with sufficiently large birefringence, calcite crystal is adopted for its high birefringence index (0.167 at 800\,\emph{nm}), and the length is chosen to be 8.98\,\emph{mm} with the optical axis parallel to incident plane, such that the time shift is designed to be 5\,\emph{ps} for one-step quantum walk.

\emph{Heralded single photon adopted as the walker --- }
The time multiplexing protocol requires pulse photons, which can be obtained by attenuating a pulse laser or modulating a continuous laser with an optical chopper. Considering the tradeoff between the operation on the time bins and the final analysis in the time domain, the time duration covers a range from tens to thousands picoseconds, reaching even a few microseconds. In our experiment, for adopting a genuine single photon as the walker and considering that the length of the crystals for realizing the time shifter should be as short as possible to reduce dispersion and improve stability, the pulse duration of the single photon should be as small as possible; it is selected on the level of hundreds of femtoseconds here. Such a short single photon pulse can be generated from SPDC with an ultrashort femtosecond pulse laser as the pump. The generated photon pairs are time correlated, and as a result, the click of detection on the idler photon can predict the existence of the signal photon. Various architectures exist for generating this type of heralded single photons from SPDC. Here, considering the features of high brightness and collection efficiency, we adopt the beamlike SPDC. Then, the length of the nonlinear crystal can be chosen to be short.

\emph{Frequency up conversion single photon detection --- }
The spectrum of the arriving time of single photons is usually obtained by the technology of time correlated single photon counting based on commercial single photon detectors. However, in our case, the signals are contained in a single photon pulse train with a pulse duration of approximately 1\,\emph{ps} and a repetition of 5\,\emph{ps}. Counting and analysis such ultra-fast single photon signals is challenging. The time resolution of commercial single photon detectors is limited by the time jitter, which is typically in the range of tens to hundreds picoseconds. As a result, it is unsuitable to directly use any commercial single photon detectors in this experiment. The detection of single photons with high resolution in time can be realized by transforming the temporal resolution to a spatial resolution. Based on the optical parameter up conversion, the measurement of an ultra-fast pulse of single photons can be realized by optical autocorrelation. That is, using an ultra-fast laser pulse to pump a nonlinear crystal, when the single photon and laser pulse meet each other inside the crystal, the single photon will be up-converted to have a short wavelength for the sum frequency process. For the photons with a long wavelength can be converted to a short one, this technology has been widely used in quantum communication for improving the detection efficiency in the infrared waveband. Here, we adopt this technology for its high resolution in time. Although periodically poled crystals are widely used in this technology for their high conversion efficiency, they are useless here for concentrating on the time resolution. The thickness of nonlinear crystal should be as thin as possible meanwhile taking into account the conversion efficiency. There exist two types of structures, collinear and non-collinear sum frequency. We adopt the latter to obtain a better signal to noise ratio (SNR), induced by the spatial divergence between the sum frequency signal and the pump laser. In this work, the crystal used is a 1\,\emph{mm} thick $\beta$-BaB$_2$O$_4$ (BBO) crystal, cut for type-\uppercase\expandafter{\romannumeral2} second harmonic generation in a beam like form. Then, the incidence angles of the signal pulse train with single photons and the pump laser are equal to each other, with $3^\circ$ to the normal direction. For reducing the noise induced by the strong pump laser, a dispersion prism in a 4F system is adopted as a spectrum filter. The scattered photons with wavelength longer than 395\,\emph{nm} are blocked by a knife edge. The rising edge in the sideband of this self-established spectrum filter is less than 1\,\emph{nm}.

\begin{figure}[h]
    \centering
    \includegraphics[width=0.45\textwidth]{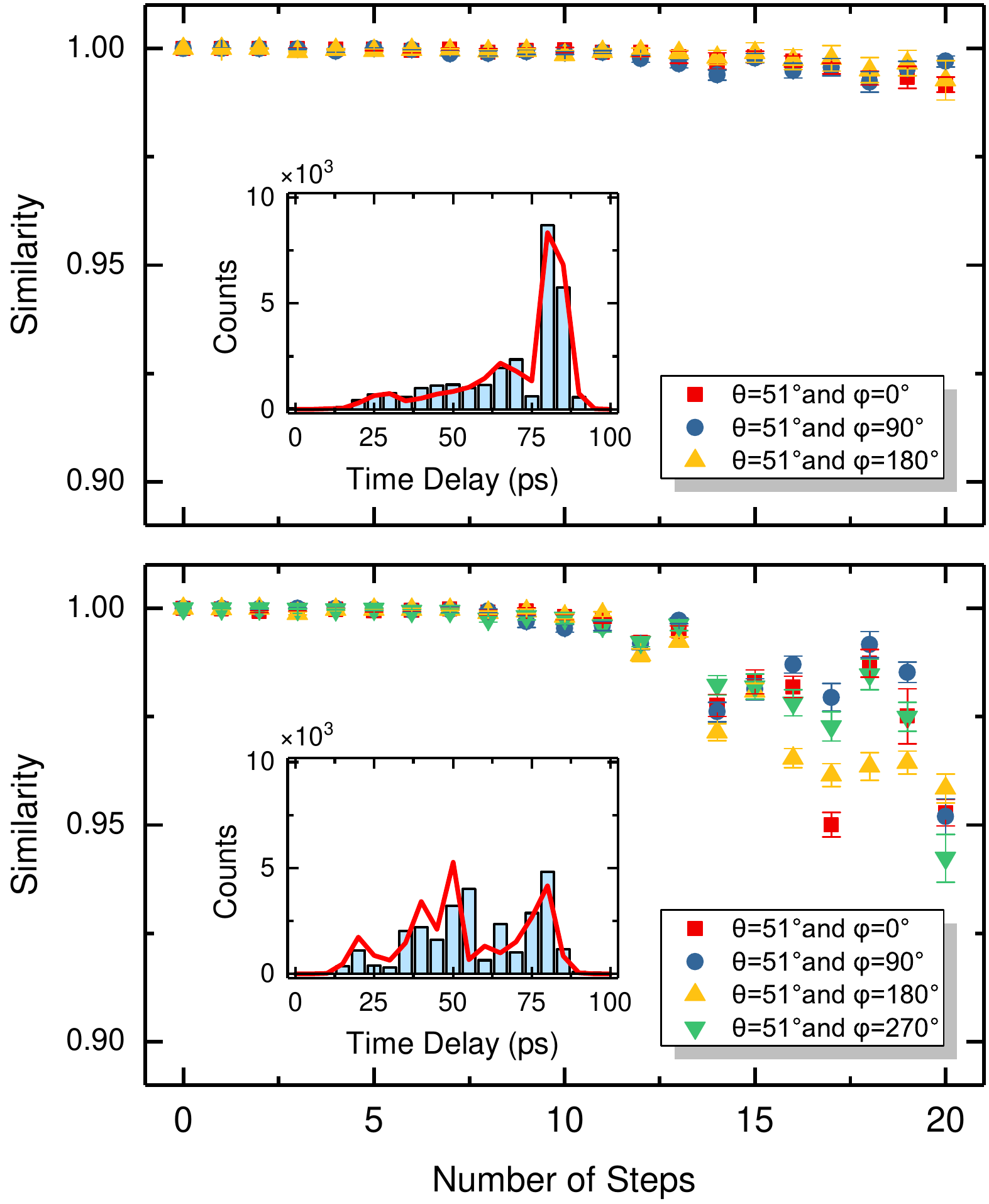}
    \caption{Plot of the similarity for each step of QW. The similarity is defined as $S=\sum_x\sqrt{P_{\text{exp}}(x)P_{\text{th}}(x)}$, where $P_{\text{exp}}$ stands for the experimental probability distribution and $P_{\text{th}}$ is the corresponding theoretical prediction. The results for ordered QW are shown in the top panel and the results for dynamically disordered QW are presented in the bottom panel. The inset in each panel shows the coincidence counts after a 20-step walk for a initial state with $\theta = 51^\circ$ and $\phi = 0^\circ$. 
    In the ordered scenarios, the similarity becomes degenerate gradually as the number of steps increases, mainly for the decoherence. In the disordered scenarios, there are more occupations near the origin (in the middle of time delays), where the interference is more complicated, results the faster degeneration of the similarity compared to the ordered cases. Only statistical errors are considered with total counts 24 thousand in 4 hundred seconds.}
    \label{fig:similarity}
\end{figure}

\emph{Description of the experimental setup --- }
An ultra fast pulse (140\,\emph{fs}) train generated by a mode-locked Ti:sapphire laser with a central wavelength at 800\,\emph{nm} and repetition ratio 76\,\emph{MHz} is firstly focused by lens L1 to shine on a 2\,\emph{mm} thick $\beta$-BaB$_2$O$_4$ crystal\,(BBO1), cut for type-\uppercase\expandafter{\romannumeral1} second harmonic generation. The frequency-doubled ultraviolet pulse (with a wavelength centered at 400\,\emph{nm}, 100\,\emph{mW} average power and horizontally polarized) and the residual pump laser are collimated by lens L2, and then separated by a dichroic mirror (DM). The frequency-doubled pulse train is then focused by lens L3 to pump the second nonlinear crystal\,(BBO2), cut for type-\uppercase\expandafter{\romannumeral2} nondegenerate beam like SPDC. The signal and idler photons are collimated together with one lens L4 (f=150\,\emph{mm}). The collimated signal photons in horizontal polarization with a center wavelength at 780\,\emph{nm} are then guided directly in free space to the following quantum walks device. The collimated idler photons in vertical polarization with a center wavelength of approximately 821\,\emph{nm} firstly pass through a spectrum filter with a central wavelength 820\,\emph{nm} and bandwidth 12\,\emph{nm} and then are coupled into a single-mode fibre and sent directly to a single-photon avalanche diode (SPAD) for counting in coincidence with the signal photons. The quantum walks device is composed of HWPs (QWPs) and calcite crystals, and each step contains one piece of HWP (QWP) and one piece of calcite crystal. In the experiment, we have adopted 20 such sets, with only 1/4 of them shown in the figure. The initial state is prepared by an apparatus composed of a polarized beam splitter (PBS1), a half wave plate (HWP1) and a quartz wave plate (QWP1). The residual pump in the frequency-double process is adopted as the pump in the following frequency up conversion single photon detection with the retroreflector R1 for temporal matching. After the quantum walks is finished, the signal photons are collected into a short single mode fibre (10\,\emph{cm} long) by a fibre collimator (FC1) and then guided to the polarization analyzer composed of QWP2, HWP2 and PBS2 successively. Finally, the arriving time of signal photons is measured by scanning the pump laser and detecting the up conversion signals with a photomultiplier tubes (PMT). For reducing the scattering noise, BBO3 is cut for noncollinear up conversion and a spectrum filter based on a 4F system is constructed, where a prism is adopted for introducing the dispersion, a knife edge is used to block the long waves and the signal is reflected to the PMT with a pickup mirror.

\section{Lempel-Ziv complexity}

Lempel-Ziv (LZ) complexity can be introduced to estimate the randomness of finite sequences, in the spirit of the Kolmogorov complexity. The LZ complexity measure counts the number of distinct substrings (patterns) in a sequence when scanned from left to right and then parsed. Note that we only consider binary sources throughout this paper. The algorithm is as follows\,\cite{hu2006}:

\begin{enumerate}
  \item Let $S_C=s_{1}s_{2}\cdots s_{n}$ denote a finite 0-1 symbolic sequence; $S_C(i,j)$ denotes a substring of $S_C$ that starts at position $i$ and ends at position $j$, that is, when $i\leqslant j$, $S_C(i,j)=s_{i}s_{i+1}\cdots s_{j}$ and when $i\geqslant j$, $S_C(i,j)=\{ \}$ (null set); $V(S_C)$ denotes the vocabulary of a sequence $S_C$. It is the set of all substrings $S_C(i,j)$ of $S_C$, (i.e., $S_C(i,j)$ for $i=1,2\cdots n$; $i\leqslant j$). For example, let $S_C=010$, we then have $V(S_C)=\{0,1,01,10,010\}$.
  \item The parsing procedure needs to scan the sequence $S_C$ from left to right. If $S_C(i,j)$ belongs to $V(S_C(1,j-1))$, then $S_C(i,j)$ and $V(S_C(1,j-1))$ is renewed to be $S_C(i,j+1)$ and $V(S_C(1,j))$, respectively. Repeat the previous process until the renewed $S_C(i,j)$ does not belong to the renewed $V(S_C(1,j-1))$, then place a dot after the renewed $S_C(i,j)$ to indicate the end of a new sequence. Update $S_C(i,j)$ and $V(S_C(1,j-1))$ to $S_C(j+1,j+1)$(the single symbol in the $j+1$ position) and $V(S_C(1,j))$, respectively, and the step 2 continues.
  \item This parsing operation begins with $S_C(1,1)$ and continues until $j=n$, where $n$ is the length of the symbolic sequence $S_C$.
\end{enumerate}

For instance, the sequence which is only composed of units 01 (i.e., $010101\cdots$) is parsed as $0\cdot 1\cdot 0101\cdots$. So the number of distinct substrings $c(n)$ is 3. To compute the LZ complexity, the sequence of $H$ and $F$ should be transformed into a 0-1 symbolic sequence (i.e., map the $H$ and $F$ to 1 and 0). Follow the above method, the twelve randomly selected sequences in Fig.4 (from 1 to 12) are parsed as:
\begin{enumerate}
  \item $1\cdot0\cdot101010101010101010\cdot\rightarrow c(n)=3$
  \item $1\cdot1110\cdot111101111011110\cdot\rightarrow c(n)=3$
  \item $1\cdot1111110\cdot0001\cdot111101\cdot11\cdot\rightarrow c(n)=5$
  \item $1\cdot0\cdot11\cdot010\cdot00\cdot111\cdot1001\cdot1010\cdot\rightarrow c(n)=8$
  \item $1\cdot110\cdot01\cdot101\cdot000\cdot111000\cdot11\cdot\rightarrow c(n)=7$
  \item $1\cdot0\cdot01\cdot110\cdot11100\cdot1010\cdot0110\cdot\rightarrow c(n)=7$
  \item $0\cdot01\cdot10\cdot0010\cdot0101\cdot1011011\cdot\rightarrow c(n)=6$
  \item $1\cdot10\cdot001\cdot111110\cdot010\cdot11101\cdot\rightarrow c(n)=6$
  \item $1\cdot11110\cdot01\cdot10010\cdot11101\cdot00\cdot\rightarrow c(n)=6$
  \item $1\cdot10\cdot001\cdot011\cdot0000\cdot1010\cdot101\cdot\rightarrow c(n)=7$
  \item $1\cdot10\cdot001\cdot1110\cdot0010\cdot0101\cdot11\cdot\rightarrow c(n)=7$
  \item $0\cdot01\cdot01011\cdot000\cdot001011\cdot111\cdot\rightarrow c(n)=6$
\end{enumerate}

\section{Detail forms of the sequences in Fig.5}

\begin{enumerate}
\item $HFHFHFHFHFHFHFHFHFHF$
\item $HHHHFHHHHFHHHHFHHHHF$
\item $HHHHHHHFFFFHHHHHFHHH$
\item $HFHHFHFFFHHHHFFHHFHF$
\item $HHHFFHHFHFFFHHHFFFHH$
\item $HFFHHHFHHHFFHFHFFHHF$
\item $FFHHFFFHFFHFHHFHHFHH$
\item $HHFFFHHHHHHFFHFHHHFH$
\item $HHHHHFFHHFFHFHHHFHFF$
\item $HHFFFHFHHFFFFHFHFHFH$
\item $HHFFFHHHHFFFHFFHFHHH$
\item $FFHFHFHHFFFFFHFHHHHH$
\end{enumerate}

\end{document}